\documentclass[11pt,twoside]{amsart}

\numberwithin{equation}{section}
\usepackage[table]{xcolor}
\usepackage{latexsym,amssymb,amsmath,amsthm,amsfonts}
\usepackage[cmtip,arrow]{xy}
\usepackage{pb-diagram,pb-xy}
\usepackage{graphicx}
\usepackage{tikz}
\usepackage{multirow}
\usepackage{enumerate}
\usetikzlibrary{matrix,arrows}
\definecolor{VerdeOlivo}{rgb}{0.3,0.5,0.1}
\definecolor{Magenta}{rgb}{.65,0.15,.2}
\definecolor{Gris}{gray}{0.7}

\newtheorem{Theorem}{Theorem}[section]
\newtheorem{Definition}[Theorem]{Definition}
\newtheorem{Lemma}[Theorem]{Lemma}

\newtheorem{Example}[Theorem]{Example}
\newtheorem{Algorithm}[Theorem]{Algorithm}
\newtheorem{Proposition}[Theorem]{Proposition}

\theoremstyle{definition}

\begin{document}


\title{Dimension Reduction in Principal Component Analysis for Trees}


\author{Carlos A.  Alfaro}
\address{
Departamento de
Matem\'aticas\\
Centro de Investigaci\'on y de Estudios Avanzados del
IPN\\
Apartado Postal
14--740 \\
07000 Mexico City, D.F.
}
\email{alfaromontufar@gmail.com}
\thanks{The first and last authors authors are partially supported by CONACyT PROINNOVA project 155874.
Also the first author was partially supported by CONACyT and the last author was partially supported by SNI.}

\author{Burcu Ayd{\i}n}
\address{
HP Laboratories\\
1501 Page Mill Rd MS 1140\\
Palo Alto, CA
}
\email{aydin@hp.com}


\author{Elizabeth Bullitt}
\address{
Department of Neurosurgery\\
University of North Carolina at Chapel Hill\\
Chapel Hill, NC
}
\email{bullitt@med.unc.edu}

\author{Alim Ladha}
\address{
Department of Neurosurgery\\
University of North Carolina at Chapel Hill\\
Chapel Hill, NC
}
\email{alim.ladha@gmail.com}

\author{Carlos E. Valencia}
\address{
Departamento de
Matem\'aticas\\
Centro de Investigaci\'on y de Estudios Avanzados del
IPN\\
Apartado Postal
14--740 \\
07000 Mexico City, D.F.
}
\email{cvalencia75@gmail.com}



\maketitle

\begin{abstract}
The statistical analysis of tree structured data is a new topic in statistics with wide application areas. Some Principal Component Analysis (PCA) ideas were previously developed for binary tree spaces. In this study, we extend these ideas to the more general space of rooted and labeled trees. We re-define concepts such as tree-line and \emph{forward} principal component tree-line for this more general space, and generalize the optimal algorithm that finds them.

We then develop an analog of classical dimension reduction technique in PCA for the tree space. To do this, we define the components that carry the least amount of variation of a tree data set, called \emph{backward} principal components. We present an optimal algorithm to find them. Furthermore, we investigate the relationship of these the forward principal components, and prove a path-independency property between the forward and backward techniques.

We apply our methods to a data set of brain artery data set of $98$ subjects. Using our techniques, we investigate how aging affects the brain artery structure of males and females. We also analyze a data set of organization structure of a large US company and explore the structural differences across different types of departments within the company.
\end{abstract}

\section{Introduction}

In statistics, data sets that reside in high dimensional spaces are quite common.
A widely used set of techniques to simplify and analyze such sets is \emph{principal component analysis} (PCA).
It was introduced by Pearson in 1901 and independently by Hotelling in 1933. A comprehensive introduction can be found in  Jolliffe (2002). 
The main aim of PCA is to provide a smaller subspace such that the maximum amount of information is retained when the original data points are projected onto it.
This smaller subspace is expressed through components. In many contexts, one dimensional subspaces are called lines, so we will follow this terminology. The line that carries the most variation present in the data set is called  \emph{first principal component} (PC1).
The \emph{second principal component} (PC2) is the line such that when combined with PC1, the most variation that can be retained in a two-dimensional subspace is kept. One may repeat this procedure to find as many principal components as necessary to properly summarize the data set in a manageable sized subspace formed by the principal components.

Another way to characterize the principal components to consider the distances of the data points to a given subspace. The line which minimizes the sum of squared distances of data points onto it can be considered as PC1. Similarly, PC2 is the line that, when combined with PC1, the sum of squared distances of the data points to this combination is minimum.

An important topic within PCA is called \emph{dimension reduction} (See Mardia et al (1973) for dimension reduction and Jolliffe (2002) pp. 144, for {\it backward elimination method}). The aim of dimension reduction method is defined as to find the components such that when eliminated, the remaining subspace will retain the maximum amount of variation. Or alternatively, the remaining subspace will have the minimum sum of squared distances to the data points. These are the components with least influence.

We would like to note that, in the general sense, any PCA method can be regarded as a dimension reduction process. However, Mardia et al (1973) reserves the term dimension reduction specifically for this method, which some other resources also refer as backward elimination, or backward PCA. In this paper we will follow Mardia et al (1973)'s convention, together with ``backward PCA" terminology. The original approach will be called \emph{forward PCA}.

In general, the choice of which technique to use depends on the needs of the end user: If only a few principal components with most variation in them are needed, then the forward approach is more suitable.
If the aim is to eliminate only a few least useful components, then the backward approach would be the appropriate choice.

The historically most common space used in statistics is the Euclidean space ($\mathbb{R}^n$) and the PCA ideas were first developed in this context. In $\mathbb{R}^n$, the two definitions of PC's (maximum variation and minimum distance) are equivalent, and the components are all orthogonal to each other. In Euclidean space, applying forward or backward PCA $n$ times for a data set in $\mathbb{R}^n$ would provide an orthogonal basis for the whole space.
Moreover, in this context, the set of components obtained with the backward approach is the same as the one obtained by the classical forward approach, only the order of the components is reversed.
This is a direct result of orthogonality properties in Euclidean space. This phenomenon can be referred as \emph{path independence} and it is very rare in non-Euclidean spaces. In fact, this paper may be presenting the first known example of path independence in non-Euclidean spaces.





With the advancement of technology, more and more data sets that do not fit into the Euclidean framework became available to researchers.
A major source of these has been biological sciences, collecting detailed images of their objects of interest using advanced imaging technologies.
The need to statically analyze such non-traditional data sets gave rise to many innovations in statistics area.
The type of non-traditional setting we will be focusing in this paper is sets of trees as data.
Such sets arise in many contexts, such as blood vessel trees (Aylward and Bullitt (2002)), lung airways trees (Tschirren et al. (2002)), and phylogenetic trees (Billera et al. (2001)).

A first starting point in PCA analysis for trees is Wang and Marron (2007), who attacked the problem of analyzing the brain artery structures obtained through a set of Magnetic Resonance Angiography (MRA) images.
They modeled the brain artery system of each subject as a binary tree and developed an analog of the forward PCA in the binary tree space.
They provided appropriate definitions of concepts such as distance, projection and line in binary tree space.
They gave formulations of first, second, etc. principal components for binary tree data sets based on these definitions.
This work has been the first study in adapting classical PCA ideas from Euclidean space to the new binary tree space.

The PCA formulations of Wang and Marron (2007) gave rise to interesting combinatorial optimization problems.
Ayd{\i}n et al. (2009) provided an algorithm to find the optimal principal components in binary tree space in linear time.
This development enabled a numerical analysis on a full-size data set of brain arteries, revealing a correlation between their structure and age.

In the context of PCA in non-Euclidean spaces, Jung et al. (2010) gave a backward PCA interpretation in image analysis. They focus on \emph{mildly non-Euclidean}, or manifold data, and propose the use of Principal Nested Spheres as a backward step-wise approach.

Marron et al. (2010) provided a concise overview of backward and forward PCA ideas and their applications in various non-classical contexts. They also mention the possibility of backwards PCA for trees:
``... The notion of backwards PCA can also generate new approaches to tree line PCA. In particular, following the backwards PCA principal in full suggests first optimizing over a number of lines together, and then iteratively reducing the number of lines." This quote essentially summarizes one of our goals in this paper.

In this work, our first goal is to extend the definitions and results of Wang and Marron (2007) and Ayd{\i}n et al. (2009) on forward PCA from binary tree space to the more general rooted labeled tree space. We will provide the generalized versions of some basic definitions such as distance, projection, PC, etc., and proceed with showing that the optimal algorithms provided for the limited binary tree space can be extended to the general rooted labeled tree space.

A rooted labeled tree is a tree such that there is a single node designated as a root, and each node is labeled in such a way that a \emph{correspondence} structure can be established between data trees. For example, in binary tree context, this means that the left and right child nodes of the any node are distinct from each other. 
In general, the labeling of the nodes greatly affects the statistical results obtained from any data set. For the rest of the paper, we will refer to the rooted labeled tree space as \emph{tree space}.

Next, we attack the problem of finding an analog of dimension reduction. We first provide a de\-fi\-ni\-tion for principal components with least influence (we call these \emph{backward principal components}) in tree space, and define the optimization problem to be solved to reach them. We then provide a linear time algorithm to solve this problem to optimality.

Furthermore, we prove that the set of backward principal components in  tree space is the same as the forward set, with order reversed, just like their counterparts in the classical Euclidean space.
This equivalence is significant since the same phenomenon in Euclidean space is a result of orthogonality, and the concept of orthogonality does not carry over to the tree space.
This result enables the analyst to switch between the two approaches as necessary while the results remain comparable, i.e., the components and their influence do not depend on which approach is used to find them. Therefore path independence property is valid in tree space PCA as well.

Our numerical results come from two main data sets. First one is an updated version of the brain artery data set previously used by Ayd{\i}n et al. (2009). Using our backward PCA tool, we investigate the effect of aging in brain artery structure in male and female subjects. We define two different kinds of age effect on the artery structure: overall branchyness and location-specific effects. We report that while both of these effects are strongly observed in males, they could not be observed in females. Secondly, we present a statistical analysis of the organization structure of a large US company. We present evidence on the structural differences across departments focusing on finance, marketing, sales and research.

The organization of the paper is as follows: In Section \ref{preliminaries}, we provide the definitions of concepts such as distance, projection, etc. in general tree space, together with a description of the forward approach and the algorithm to solve it.
These are generalizations of the concepts introduced in Wang and Marron (2007) and Ayd{\i}n et al (2009). In Section \ref{backward} we describe the problem of finding the backward principal components in tree space and give an algorithm to find the optimal solution.
In Section $4$ we prove the equivalence of forward and backward approaches in tree space. Section \ref{numerical} contains our numerical analysis results.



\section{Forward PCA in Tree Space}\label{preliminaries}

In this section, we will provide definitions of some key concepts such as distance, projection, etc. in tree space, together with illustrative examples. The binary tree space versions of these definitions were previously given in Wang and Marron (2007) and Ayd{\i}n et al. (2009). We will also provide the tree space versions of their PCA results, and prove their optimality in the more general tree space.

In this paper the term \emph{tree} is reserved for rooted tree graphs in which each node is distinguished from each other through labels. The labeling method can differ depending on the properties of any tree data set. For labeling binary trees, Wang and Marron (2007) uses a level-order indexing method. In this scheme the root node has index 1.
For the remaining nodes, if a node has index $i$, then the index of its left child is $2i$ and of its right child is $2i+1$. (See Figure $1$). Labeling general trees may get significantly more complicated.
\begin{center}
	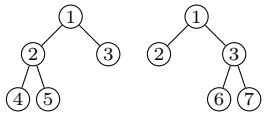
\begin{figure}[h]\label{Figure1}
		\[
		\begin{array}{cc}
			\begin{tikzpicture}[scale=1]
				\tikzstyle{every node}=[draw, circle, inner sep=1pt]
				\draw (0,0) node (r) {\tiny 1};
				\draw (-.5,-.5) node (v1) {\tiny 2};
				\draw (-.7,-1.1) node (v2) {\tiny 4};
				\draw (-.3,-1.1) node (v3) {\tiny 5};
				\draw (.5,-.5) node (v4) {\tiny 3};
				\draw (v4) -- (r) -- (v1) -- (v2)
					(v3) -- (v1);
			\end{tikzpicture}
		&
			\begin{tikzpicture}[scale=1]
				\tikzstyle{every node}=[draw, circle, inner sep=1pt]
				\draw (0,0) node (r) {\tiny 1};
				\draw (-.5,-.5) node (v1) {\tiny 2};
				\draw (.5,-.5) node (v4) {\tiny 3};
				\draw (.7,-1.1) node (v5) {\tiny 7};
				\draw (.3,-1.1) node (v6) {\tiny 6};
				\draw (v5) -- (v4) -- (r) -- (v1)
					(v4) -- (v6);
			\end{tikzpicture}
		\end{array}
		\]
		\caption{Two trees of which nodes are labeled using level-order indexing method. The children of any node are distinct from each other. The nodes 1,2 and 3 in the left data tree correspond to the nodes 1,2 and 3 in the right data tree.}
	\end{figure}
\end{center}

A data set, $\mathcal{T}$, is an indexed finite set of $n$ trees.
A distance metric between two trees is the symmetric difference of their nodes. Given two trees, $t_1$ and $t_2$, the {\bf distance} between $t_1$ and $t_2$, denoted by $d(t_1,t_2)$, is
\[
	|t_1\setminus t_2|+|t_2\setminus t_1|,
\]
where $|\cdot|$ is the number of nodes and $\setminus$ is the node set difference. In Figure $1$, the nodes 1, 2 and 3 are common to both of the trees, so they do not contribute to the distance between them. The nodes 4,5, 6 and 7 exist in one data tree but not in the other, therefore, the distance between the left and right trees in the figure is $|\{4,5,6,7\}|=4$.

The {\bf support tree} and the {\bf intersection tree} of a data set $\mathcal{T}=\{t_1, \dots, t_n\}$ are defined as:
\[
	Supp(\mathcal{T})= \cup_{i=1}^{n}t_i \text{ and } Int(\mathcal{T})= \cap_{i=1}^{n}t_i,
\]
respectively.

As before, the line concept is a close counterpart to the lines in Euclidean space. In the most general sense line refers to a set of points that are next to each other. These points lie in a given direction, which makes the line ``one-dimensional". Due to the discrete nature of tree space, the points (trees) that are next to each other are defined the points with distance $1$, the smallest possible distance between two non-identical trees. To mimic the one-dimensional direction property, we require that every next point on the line in tree space is obtained by adding a child of most recently added node. The resulting construct is a set of trees that start from a starting tree and expands following a path away from the root, which is akin to the sense of direction in Euclidean space. A formal definition of a line in tree space is given as follows:
\begin{Definition}
	Given a data set $\mathcal{T}$, a {\bf tree-line}, ${L=\{ l_0, \dots, l_k\}}$, is a sequence of trees where $l_0$ is called the starting tree, and $l_{i}$ is defined from $l_{i-1}$ by the addition of a single node $v_i\in Supp(\mathcal{T})$.
	In addition, each $v_{i}$ is a child of $v_{i-1}$.
\end{Definition}
See Example \ref{exa:exa1} for an example tree-line.

The next concept to construct is the projection in this space. In general, the projection of a point onto an object can be defined as the closest point on the object to the projected point. This can be formalized in tree space as:
\begin{Definition}
	The {\bf projection} of a tree $t$ onto the tree-line $L$ is
	\[
		P_L(t)= \arg \min_{\tiny l\in L} \{d(t,l) \}
	\]
\end{Definition}
The projection of a data tree onto a tree-line can be regarded as the point in the tree-line most similar to the data tree.

Example \ref{exa:exa1} contains a small data set and a tree-line, and illustrates how the projection of each data point onto the given tree-line can be found.

\begin{Example}\label{exa:exa1}
Let us consider the following data set consisting of $3$ data points. For simplicity, we use a set consisting of binary trees only.

\[
\mathcal{T}
=\left\{
\begin{array}{ccccc}
	t_1=
	\begin{tikzpicture}[scale=1]
		\tikzstyle{every node}=[draw, fill, circle, inner sep=1pt]
		\draw (0,0) node (r) {};
		\draw (-.6,-.5) node (v1) {};
		\draw (-.8,-1) node (v11) {};
		\draw (-.4,-1) node (v12) {};
		\draw (0,-.5) node (v2) {};
		\draw (.2,-1) node (v22) {};
		\draw (v1) -- (r) -- (v2)
			  (v11) -- (v1) -- (v12)
			  (v2) --(v22);
	\end{tikzpicture}
	&
	,
	&
	t_2=
	\begin{tikzpicture}[scale=1]
		\tikzstyle{every node}=[draw, fill, circle, inner sep=1pt]
		\draw (0,0) node (r) {};
		\draw (-.6,-.5) node (v1) {};
		\draw (-.4,-1) node (v12) {};
		\draw (0,-.5) node (v2) {};
		\draw (-.2,-1) node (v21) {};
		\draw (.6,-.5) node (v3) {};
		\draw (v1) -- (r) -- (v2)
			  (v1) -- (v12)
			  (v21) -- (v2)
			  (r) -- (v3);
	\end{tikzpicture}
	&
	,
	&
	t_3=
	\begin{tikzpicture}[scale=1]
		\tikzstyle{every node}=[draw, fill, circle, inner sep=1pt]
		\draw (0,0) node (r) {};
		\draw (0,-.5) node (v2) {};
		\draw (-.2,-1) node (v21) {};
		\draw (.2,-1) node (v22) {};
		\draw (.6,-.5) node (v3) {};
		\draw (.4,-1) node (v31) {};
		\draw (.8,-1) node (v32) {};
		\draw (r) -- (v2)
			  (v21) -- (v2) --(v22)
			  (r) -- (v3)
			  (v31) -- (v3) -- (v32);
	\end{tikzpicture}
\end{array}
\right\},
\]

\[
	Supp(\mathcal{T})
	=
	\begin{tikzpicture}[scale=1]
		\tikzstyle{every node}=[draw, fill, circle, inner sep=1pt]
		\draw (0,0) node (r) {};
		\draw (-.6,-.5) node (v1) {};
		\draw (-.8,-1) node (v11) {};
		\draw (-.4,-1) node (v12) {};
		\draw (0,-.5) node (v2) {};
		\draw (-.2,-1) node (v21) {};
		\draw (.2,-1) node (v22) {};
		\draw (.6,-.5) node (v3) {};
		\draw (.4,-1) node (v31) {};
		\draw (.8,-1) node (v32) {};
		\draw (v1) -- (r) -- (v2)
			  (v11) -- (v1) -- (v12)
			  (v21) -- (v2) --(v22)
			  (r) -- (v3)
			  (v31) -- (v3) -- (v32);
	\end{tikzpicture}
\]
and a tree-line

\[
L
=\left\{
\begin{array}{ccccccc}
	l_0=
	\begin{tikzpicture}[scale=1]
		\tikzstyle{every node}=[draw, fill, circle, inner sep=1pt]
		\draw (0,0) node (r) {};
		\draw (-.6,-.5) node (v1) {};
		\draw (-.8,-1) node (v11) {};
		\draw (v1) -- (r)
			  (v11) -- (v1);
	\end{tikzpicture}
	&
	,
	&
	l_1=
	\begin{tikzpicture}[scale=1]
		\tikzstyle{every node}=[draw, fill, circle, inner sep=1pt]
		\draw (0,0) node (r) {};
		\draw (-.6,-.5) node (v1) {};
		\draw (-.8,-1) node (v11) {};
		\draw (0,-.5) node (v2) {};
		\draw (v1) -- (r) -- (v2)
			  (v11) -- (v1);
	\end{tikzpicture}
	&
	,
	&
	l_2=
	\begin{tikzpicture}[scale=1]
		\tikzstyle{every node}=[draw, fill, circle, inner sep=1pt]
		\draw (0,0) node (r) {};
		\draw (-.6,-.5) node (v1) {};
		\draw (-.8,-1) node (v11) {};
		\draw (0,-.5) node (v2) {};
		\draw (.2,-1) node (v22) {};
		\draw (v1) -- (r) -- (v2)
			  (v11) -- (v1)
			  (v2) --(v22);
	\end{tikzpicture}
\end{array}
\right\}.
\]
The following table gives the distance between each tree of $\mathcal{T}$ and each tree of $L$:

\begin{center}
	\begin{tabular}{|c|ccc|}
		\hline & $l_0$ & $l_1$ & $l_2$ \\
		\hline $t_1$ & 3 & 2 & 1 \\
		\hline $t_2$ & 5 & 4 & 5 \\
		\hline $t_3$ & 8 & 7 & 6 \\
		\hline
	\end{tabular}
\end{center}
So, we can observe that $P_L(t_1)=l_2$,  $P_L(t_2)=l_1$ and  $P_L(t_3)=l_2$.
\end{Example}

Finally, we will define the concept of ``path" that will be useful later on.
\begin{Definition}
	Given a tree-line $L=\{l_0, \cdots, l_k \}$, the {\bf path} of $L$ is the unique path from the root to $v_k$, the last node added in L, and it is denoted by $p_L$.
\end{Definition}
Note that our path definition is different than the one given in Ayd{\i}n et al. (2009), which included only the nodes added to the starting tree instead of forming a set starting from the root node.

The next lemma provides an easy-to-use a formula for the projection of a data point. The proof of it can be found in the Appendix.
\begin{Lemma}\label{lemma1}
	Let $t$ be a binary tree and $L=\{l_0, \cdots, l_k \}$ be a tree-line.
	Then
	\[
		P_L(t)=l_0\cup( t \cap p_L).
	\]
\end{Lemma}
It follows that projection of a tree over a tree-line is unique.

Wang and Marron (2007) gave a definition of first principal component tree-line in the binary tree space. It was defined as the tree-line that minimizes the sum of distances of the data points to their projections on the line. This can be viewed as the one-dimensional line that best fits the data. We will provide their definition below, adopted to the general tree space. We also note that this is the ``forward PCA" approach where a subspace that carries the most amount of variation is sought. We will develop the ``backward PCA" approach in the upcoming section.

\begin{Definition}
	For a data set $\mathcal{T}$ and the set of all tree-lines $\mathcal{L}$ in $Supp(\mathcal{T})$ with the same starting point $l_0$, the {\bf first (forward) principal component tree-line}, PC1, is
	\[
		L_1^f=\arg \min_{L\in \mathcal{L}} \sum_{t\in \mathcal{T}} d(t,P_L(t)).
	\]
\end{Definition}

As we will see in Example \ref{exa:forward}, the definition of the principal components allows multiple solutions.
A tie-breaking rule depending on the nature of the data should be established to reach consistent results in the existence of ties.
In order to have a tie breaking rule dealing with the  PC's definition, we assume that the set of all tree-lines is totally ordered.
This tie-breaking rule (total order) can be induced to the set of paths.
Thus, we denote by $p_L>p_{L'}$ that the path $p_L$ is preferred to $p_{L'}$.

For an analogous notion of the additional components in tree space, we need to define the concept of the union of tree-lines, and projection onto a union. We say that given tree-lines $L_1=\{l_{1,0}, l_{1,1}, \dots, l_{1,m_1}\}$, \dots, $L_q=\{l_{q,0}, l_{m,1}, \dots, l_{q,m_q}\}$, their {\bf union} is the set of all possible unions of members of $L_1$ trough $L_q$:
	\begin{eqnarray*}
		L_1\cup\cdots \cup L_q & = & \{l_{1,i_1}\cup\cdots\cup l_{m,i_m} \mid i_1\in \{1,\cdots, m_1\}, \cdots, i_q\in \{0, \cdots, m_q\} \}.
	\end{eqnarray*}
In light of this, the projection of a tree $t$ onto $L_1 \cup \cdots \cup L_q$ is:
	\[
		P_{L_1 \cup \cdots \cup L_q}(t)=\arg \min_{\tiny l\in L_1 \cup \cdots \cup L_q} \{d(t,l) \}
	\]
Next, we provide the definition of the general $k^{th}$ PC:
\begin{Definition}
	For a data set $\mathcal{T}$ and the set of all tree-lines $\mathcal{L}$ in $Supp(\mathcal{T})$ with the same starting point $l_0$, the {\bf $k$-th (forward) principal component tree-line}, PCk, is defined recursively as
	\[
		L_k^f=\arg \min_{L\in \mathcal{L}} \sum_{t\in \mathcal{T}} d(t,P_{L_1^f \cup \cdots \cup L_{k-1}^f\cup L}(t)).
	\]
The path of the $k$-th principal component tree-line will be denoted by $p_k^f$.
\end{Definition}
The following lemma describes a key property that will be used to interpret the projection of a tree onto a subspace defined by a set of tree-lines. The reader may refer to the Appendix for the proof.
\begin{Lemma}\label{lemma2}
	Let $L_1, L_2, \dots, L_q$ be tree-lines with a common starting point, and $t$ be a tree.
	Then
	\[
		P_{L_1 \cup \cdots \cup L_q}(t)=P_{L_1}(t) \cup \cdots \cup P_{L_q}(t)
	\]
\end{Lemma}
Ayd{\i}n et al. (2009) provided a linear time algorithm to find the forward principal components in binary tree space. We will give a generalization of that algorithm in tree space, and prove that the extended version also gives the optimal PC's. The algorithm uses the weight function $w_k(v)$, defined as follows:
\begin{Definition}
	Let $\mathcal{T}$ be a data set and $\mathcal{L}$ be the set of all tree-lines with the same starting point $l_0$. Let $\delta$ be an indicator function, defined as $\delta(v,t)=1$ if $v\in t$, and $0$ otherwise.
	Given $L_1^f, \dots, L_{k-1}^f$, the first $k-1$ PC tree-lines.
	The $k$-th weight of a node $v\in Supp(\mathcal{T})$ is
	\[
		w_k(v)=
		\begin{cases}
			0, & \text{ if } v\in l_0\cup p_1^f \cup \cdots \cup p_{k-1}^f,\\
			\sum_{t\in \mathcal{T}}\delta(v,t), & otherwise.
		\end{cases}
	\]
\end{Definition}

The following algorithm computes the $k$-th PC tree-line:

\begin{Algorithm}{Forward algorithm.}
	Let $\mathcal{T}$ be a data set and $\mathcal{L}$ be the set of all tree-lines with the same starting point $l_0$.\\
	{\bf Input:} $L_1^f, \dots, L_{k-1}^f$, the first $(k-1)$-st PC tree-lines.\\
	{\bf Output:} A tree-line.\\
	Return the tree-line whose path maximizes the sum of $w_k$ weights in the support tree. Break ties according to an appropriate tie-breaking rule.
\end{Algorithm}

To better explain how the algorithm works, we will apply the forward algorithm to the toy data set given in Example \ref{exa:exa1}.

\begin{Example}\label{exa:forward}
In this example, we select as tie-breaking rule the tree-line with leftmost path.
We take the intersection tree as the starting point (illustrated in red below). The table given below summarizes iterations of the algorithm, where each row corresponds to one iteration. At each of the iterations, the name of the principal component obtained at that iteration is given in left column. The support tree with updated weights ($w_i'(.)$) is given in the middle column. The paths of selected PC tree-lines according to these weights is given in right column.
\begin{center}
\begin{tabular}{ccc}
PC 1 &
	\begin{tikzpicture}[scale=1]
		\tikzstyle{every node}=[draw, fill, circle, inner sep=1pt]
		\draw (0,0) node[label=above:{\tiny $0$}, color=red] (r) {};
		\draw (-.6,-.5) node[label=left:{\tiny $2$}] (v1) {};
		\draw (-.8,-1) node[label=below:{\tiny $1$}] (v11) {};
		\draw (-.4,-1) node[label=below:{\tiny $2$}] (v12) {};
		\draw (0,-.5) node[label=left:{\tiny $0$}, color=red] (v2) {};
		\draw (-.2,-1) node[label=below:{\tiny $2$}] (v21) {};
		\draw (.2,-1) node[label=below:{\tiny $2$}] (v22) {};
		\draw (.6,-.5) node[label=right:{\tiny $2$}] (v3) {};
		\draw (.4,-1) node[label=below:{\tiny $1$}] (v31) {};
		\draw (.8,-1) node[label=below:{\tiny $1$}] (v32) {};
		\draw (v1) -- (r)
			  (v11) -- (v1) -- (v12)
			  (v21) -- (v2) -- (v22);
		\draw[thick, color=red] (r) -- (v2);
		\draw (r) -- (v3)
			  (v31) -- (v3) -- (v32);
	\end{tikzpicture}
	&
	\begin{tikzpicture}[scale=1]
		\tikzstyle{every node}=[draw, fill, circle, inner sep=1pt]
		\draw (0,0) node[label=above:{\tiny $0$}, color=red] (r) {};
		\draw (-.6,-.5) node[label=left:{\tiny $2$}] (v1) {};
		\draw (-.4,-1) node[label=below:{\tiny $2$}] (v12) {};
		\draw (v1) -- (r)
			  (v12) -- (v1);
	\end{tikzpicture}

\\
PC 2 &
	\begin{tikzpicture}[scale=1]
		\tikzstyle{every node}=[draw, fill, circle, inner sep=1pt]
		\draw (0,0) node[label=above:{\tiny $0$}, color=red] (r) {};
		\draw (-.6,-.5) node[label=left:{\tiny $0$}] (v1) {};
		\draw (-.8,-1) node[label=below:{\tiny $1$}] (v11) {};
		\draw (-.4,-1) node[label=below:{\tiny $0$}] (v12) {};
		\draw (0,-.5) node[label=left:{\tiny $0$}, color=red] (v2) {};
		\draw (-.2,-1) node[label=below:{\tiny $2$}] (v21) {};
		\draw (.2,-1) node[label=below:{\tiny $2$}] (v22) {};
		\draw (.6,-.5) node[label=right:{\tiny $2$}] (v3) {};
		\draw (.4,-1) node[label=below:{\tiny $1$}] (v31) {};
		\draw (.8,-1) node[label=below:{\tiny $1$}] (v32) {};
		\draw (v1) -- (r)
			  (v11) -- (v1) -- (v12)
			  (v21) -- (v2) -- (v22);
		\draw[thick, color=red] (r) -- (v2);
		\draw (r) -- (v3)
			  (v31) -- (v3) -- (v32);
	\end{tikzpicture}
	&
	\begin{tikzpicture}[scale=1]
		\tikzstyle{every node}=[draw, fill, circle, inner sep=1pt]
		\draw (0,0) node[label=above:{\tiny $0$}, color=red] (r) {};
		\draw (.6,-.5) node[label=right:{\tiny $2$}] (v3) {};
		\draw (.4,-1) node[label=below:{\tiny $1$}] (v31) {};
		\draw (r) -- (v3) -- (v31);
	\end{tikzpicture}

\\
PC 3 &
	\begin{tikzpicture}[scale=1]
		\tikzstyle{every node}=[draw, fill, circle, inner sep=1pt]
		\draw (0,0) node[label=above:{\tiny $0$}, color=red] (r) {};
		\draw (-.6,-.5) node[label=left:{\tiny $0$}] (v1) {};
		\draw (-.8,-1) node[label=below:{\tiny $1$}] (v11) {};
		\draw (-.4,-1) node[label=below:{\tiny $0$}] (v12) {};
		\draw (0,-.5) node[label=left:{\tiny $0$}, color=red] (v2) {};
		\draw (-.2,-1) node[label=below:{\tiny $2$}] (v21) {};
		\draw (.2,-1) node[label=below:{\tiny $2$}] (v22) {};
		\draw (.6,-.5) node[label=right:{\tiny $0$}] (v3) {};
		\draw (.4,-1) node[label=below:{\tiny $0$}] (v31) {};
		\draw (.8,-1) node[label=below:{\tiny $1$}] (v32) {};
		\draw (v1) -- (r)
			  (v11) -- (v1) -- (v12)
			  (v21) -- (v2) -- (v22);
		\draw[thick, color=red] (r) -- (v2);
		\draw (r) -- (v3)
			  (v31) -- (v3) -- (v32);
	\end{tikzpicture}
	&
	\begin{tikzpicture}[scale=1]
		\tikzstyle{every node}=[draw, fill, circle, inner sep=1pt]
		\draw (0,0) node[label=above:{\tiny $0$}, color=red] (r) {};
		\draw (0,-.5) node[label=left:{\tiny $0$}, color=red] (v2) {};
		\draw (-.2,-1) node[label=below:{\tiny $2$}] (v21) {};
		\draw (v21) -- (v2);
		\draw[thick, color=red] (r) -- (v2);
	\end{tikzpicture}
\\
PC 4 &
	\begin{tikzpicture}[scale=1]
		\tikzstyle{every node}=[draw, fill, circle, inner sep=1pt]
		\draw (0,0) node[label=above:{\tiny $0$}, color=red] (r) {};
		\draw (-.6,-.5) node[label=left:{\tiny $0$}] (v1) {};
		\draw (-.8,-1) node[label=below:{\tiny $1$}] (v11) {};
		\draw (-.4,-1) node[label=below:{\tiny $0$}] (v12) {};
		\draw (0,-.5) node[label=left:{\tiny $0$}, color=red] (v2) {};
		\draw (-.2,-1) node[label=below:{\tiny $0$}] (v21) {};
		\draw (.2,-1) node[label=below:{\tiny $2$}] (v22) {};
		\draw (.6,-.5) node[label=right:{\tiny $0$}] (v3) {};
		\draw (.4,-1) node[label=below:{\tiny $0$}] (v31) {};
		\draw (.8,-1) node[label=below:{\tiny $1$}] (v32) {};
		\draw (v1) -- (r)
			  (v11) -- (v1) -- (v12)
			  (v21) -- (v2) -- (v22);
		\draw[thick, color=red] (r) -- (v2);
		\draw (r) -- (v3)
			  (v31) -- (v3) -- (v32);
	\end{tikzpicture}
	&
	\begin{tikzpicture}[scale=1]
		\tikzstyle{every node}=[draw, fill, circle, inner sep=1pt]
		\draw (0,0) node[label=above:{\tiny $0$}, color=red] (r) {};
		\draw (0,-.5) node[label=left:{\tiny $0$}, color=red] (v2) {};
		\draw (.2,-1) node[label=below:{\tiny $2$}] (v22) {};
		\draw (v2) -- (v22);
		\draw[thick, color=red] (r) -- (v2);
	\end{tikzpicture}
\\
PC 5 &
	\begin{tikzpicture}[scale=1]
		\tikzstyle{every node}=[draw, fill, circle, inner sep=1pt]
		\draw (0,0) node[label=above:{\tiny $0$}, color=red] (r) {};
		\draw (-.6,-.5) node[label=left:{\tiny $0$}] (v1) {};
		\draw (-.8,-1) node[label=below:{\tiny $1$}] (v11) {};
		\draw (-.4,-1) node[label=below:{\tiny $0$}] (v12) {};
		\draw (0,-.5) node[label=left:{\tiny $0$}, color=red] (v2) {};
		\draw (-.2,-1) node[label=below:{\tiny $0$}] (v21) {};
		\draw (.2,-1) node[label=below:{\tiny $0$}] (v22) {};
		\draw (.6,-.5) node[label=right:{\tiny $0$}] (v3) {};
		\draw (.4,-1) node[label=below:{\tiny $0$}] (v31) {};
		\draw (.8,-1) node[label=below:{\tiny $1$}] (v32) {};
		\draw (v1) -- (r)
			  (v11) -- (v1) -- (v12)
			  (v21) -- (v2) -- (v22);
		\draw[thick, color=red] (r) -- (v2);
		\draw (r) -- (v3)
			  (v31) -- (v3) -- (v32);
	\end{tikzpicture}
	&
	\begin{tikzpicture}[scale=1]
		\tikzstyle{every node}=[draw, fill, circle, inner sep=1pt]
		\draw (0,0) node[label=above:{\tiny $0$}, color=red] (r) {};
		\draw (-.6,-.5) node[label=left:{\tiny $0$}] (v1) {};
		\draw (-.8,-1) node[label=below:{\tiny $1$}] (v11) {};
		\draw (v1) -- (r)
			  (v11) -- (v1);
	\end{tikzpicture}
\\
PC 6 &
	\begin{tikzpicture}[scale=1]
		\tikzstyle{every node}=[draw, fill, circle, inner sep=1pt]
		\draw (0,0) node[label=above:{\tiny $0$}, color=red] (r) {};
		\draw (-.6,-.5) node[label=left:{\tiny $0$}] (v1) {};
		\draw (-.8,-1) node[label=below:{\tiny $0$}] (v11) {};
		\draw (-.4,-1) node[label=below:{\tiny $0$}] (v12) {};
		\draw (0,-.5) node[label=left:{\tiny $0$}, color=red] (v2) {};
		\draw (-.2,-1) node[label=below:{\tiny $0$}] (v21) {};
		\draw (.2,-1) node[label=below:{\tiny $0$}] (v22) {};
		\draw (.6,-.5) node[label=right:{\tiny $0$}] (v3) {};
		\draw (.4,-1) node[label=below:{\tiny $0$}] (v31) {};
		\draw (.8,-1) node[label=below:{\tiny $1$}] (v32) {};
		\draw (v1) -- (r)
			  (v11) -- (v1) -- (v12)
			  (v21) -- (v2) -- (v22);
		\draw[thick, color=red] (r) -- (v2);
		\draw (r) -- (v3)
			  (v31) -- (v3) -- (v32);
	\end{tikzpicture}
	&
	\begin{tikzpicture}[scale=1]
		\tikzstyle{every node}=[draw, fill, circle, inner sep=1pt]
		\draw (0,0) node[label=above:{\tiny $0$}, color=red] (r) {};
		\draw (.6,-.5) node[label=right:{\tiny $0$}] (v3) {};
		\draw (.8,-1) node[label=below:{\tiny $1$}] (v32) {};
		\draw (r) -- (v3) -- (v32);
	\end{tikzpicture}
\end{tabular}
\end{center}
\end{Example}

The next theorem states that the tree-line returned by the forward algorithm is precisely the $k$-th PC tree-line. The proof is in the Appendix.
\begin{Theorem}\label{theorem1}
	Let $\mathcal{T}$ be a data set and $\mathcal{L}$ be the set of all tree-lines with the same starting point $l_0$.
	Let $L_1^f, \dots, L_{k-1}^f$ be the first $(k-1)$-st PC tree-lines.
	Then, the forward algorithm returns the $k$th PC tree-line, $L_{k}^f$.
\end{Theorem}
In theory, an arbitrary line would extend to infinity. In this paper, we limit our scope to the line pieces that reside within the support tree of a given data set since extending lines outside of support tree's scope would introduce unnecessary trivialities. Within this restriction, it can be seen that the possible principal component tree-lines for a given data set are those that theirs paths are maximum (there is no other path in $Supp(\mathcal{T})$ containing $p_L$). We also consider only the tree-lines that are not trivial (the tree-line consist of $l_0$ and at least one more point).

In the light of this, we let $\mathcal{L_P}$ denote the set of all maximal non trivial tree-lines with staring point $l_0$, contained in $Supp(\mathcal{T})$. Also we name $\mathcal{P}$ to be the set of all paths in $Supp(\mathcal{T})$ from the root to leaves that are not in $l_0$. It is easy to see that $\mathcal{P}$ is the set of paths of tree-lines in $\mathcal{L_P}$. Also note that $|\mathcal{L_P}|=|\mathcal{P}|=n$ and $\displaystyle Supp(\mathcal{T})=l_0 \cup\bigcup_{p_L\in \mathcal{P}} p_L$.

\section{Dimension Reduction for Rooted Trees}\label{backward}

In this section, we will define \emph{backward principal component tree-lines}. This structure is the tree space equivalent of the backward principal component in the classical dimension reduction setting. They represent the directions that carry the least information about the data set and thus can be taken out. Our definition describes backward principal components as directions such that when eliminated, the remaining subspace will retain the maximum amount of variation. Or alternatively, the remaining subspace will have the minimum sum of squared distances to the data points. These are considered to be the components with least influence.
We also present an algorithm that finds these components, and we provide a theoretical result proving the optimality of our algorithm.

While using the backward approach, we must use the opposite tie-breaking rule we used in the forward approach.
That is, $p_L>p_{L'}$ means that the path $p_{L'}$ is preferred to $p_{L}$.

\begin{Definition}
	For a data set $\mathcal{T}$ and the set of tree-lines $\mathcal{L_P}$ with the same starting point $l_0$, the {\bf $\bf n^{th}$ backward principal component tree-line}, \emph{BPCn}, is
	\[
		L_n^{b}=\arg \min_{L\in \mathcal{L_P}} \sum_{t\in \mathcal{T}} d(t,P_{\bigcup L'\in\mathcal{L_P}\setminus \{L\}}(t)).
	\]
	The {\bf $\bf (n-k)^{th}$ backward principal component tree-line} is defined recursively as
	\begin{eqnarray}\label{eqn:kBCP}
		L_{n-k}^{b}= & \arg \min_{L\in \mathcal{L_P}\setminus \{L_n^{b}, \cdots, L_{n-k+1}^{b}\} } \sum_{t\in \mathcal{T}} d(t,P_{\bigcup L'\in\mathcal{L_P}\setminus \{L_n^{b}, \cdots, L_{n-k+1}^{b} , L\}}(t)).
	\end{eqnarray}
\end{Definition}
The path associated to the $(n-k)$-th backward principal component tree-line will be denoted by $p_{n-k}^b$.
The following node weight definition will be key to the upcoming algorithm for finding backward components:
\begin{Definition}
	Let $\mathcal{T}$ be a data set and $\mathcal{L}$ be the set of all tree-lines with the same starting point $l_0$.
	Let $L_n^{b}, \dots, L_{n-k+1}^{b}$ be the last $k$ BPC tree-lines and $\textbf{B}=\mathcal P\setminus \{p_{n}^b,\dots , p_{n-k+1}^{b}\}$.
	For $v\in Supp({\bf B})$, the $(n-k)$-th backward weight of the node $v$ is
	\[
		w_{n-k}'(v)=
		\begin{cases}
			0 & \displaystyle \text{If } v\in l_0 \text{ or } v \text{ belongs to at least two different paths of }{\bf B}\\
			\sum_{t\in \mathcal{T}}\delta(v,t) & \text{Otherwise.}\\
		\end{cases}
	\]
\end{Definition}

The following algorithm computes the backward principal components.

\begin{Algorithm}{Backward Algorithm.}
	Let $\mathcal{T}$ be a data set of binary set and $\mathcal{L}$ be the set of all tree-lines on $Supp(\mathcal{T})$ with the same starting point $l_0$.\\
	{\bf Input:} $L_n^{b}, \dots, L_{n-k+1}^{b}$, the last $k$ BPC tree-lines.\\
	{\bf Output:} $L_{n-k}^b$, the $(n-k)^{th}$ BPC tree-line.\\
	Let $\textbf{B}=\mathcal P\setminus \{p_{n}^b,\dots , p_{n-k+1}^{b}\}$.\\
	Return the tree-line $L_{n-k}^b$ whose path minimizes the sum of $w_k'$ weights in the support tree $Supp({\bf B})$.
	If there are more than one candidate, select the tree-line according to an appropriate tie-breaking rule (it coincides with the opposite tie-breaking rule used in the forward algorithm).
\end{Algorithm}

As the forward algorithm explained in previous section, the backward algorithm also finds the optimal solution in linear time.

Next, we provide an example illustrating the steps of the backward algorithm.
We will apply the backward algorithm to the toy data set given in Example \ref{exa:exa1}.
In this example, we use the same starting point as in example \ref{exa:forward}.
Furthermore, we use the opposite tie-breaking rule we used in the forward algorithm, in this case is to select the rightmost tree-line.

\begin{Example}
The table given below summarizes iterations of the algorithm, where each row corresponds to one iteration. At each of the iterations, the name of the backward principal component obtained at that iteration is given in left column. The pruned support tree with updated weights ($w_i'(.)$) is given in the middle column. The paths of selected PC tree-lines according to these weights is given in right column.

\begin{center}
\begin{tabular}{ccc}
BPC 6 &
	\begin{tikzpicture}[scale=1]
		\tikzstyle{every node}=[draw, fill, circle, inner sep=1pt]
		\draw (0,0) node[label=above:{\tiny $0$}, color=red] (r) {};
		\draw (-.6,-.5) node[label=left:{\tiny $0$}] (v1) {};
		\draw (-.8,-1) node[label=below:{\tiny $1$}] (v11) {};
		\draw (-.4,-1) node[label=below:{\tiny $2$}] (v12) {};
		\draw (0,-.5) node[label=left:{\tiny $0$}, color=red] (v2) {};
		\draw (-.2,-1) node[label=below:{\tiny $2$}] (v21) {};
		\draw (.2,-1) node[label=below:{\tiny $2$}] (v22) {};
		\draw (.6,-.5) node[label=right:{\tiny $0$}] (v3) {};
		\draw (.4,-1) node[label=below:{\tiny $1$}] (v31) {};
		\draw (.8,-1) node[label=below:{\tiny $1$}] (v32) {};
		\draw (v1) -- (r)
			  (v11) -- (v1) -- (v12)
			  (v21) -- (v2) -- (v22);
		\draw[thick, color=red] (r) -- (v2);
		\draw (r) -- (v3)
			  (v31) -- (v3) -- (v32);
	\end{tikzpicture}
	&
	\begin{tikzpicture}[scale=1]
		\tikzstyle{every node}=[draw, fill, circle, inner sep=1pt]
		\draw (0,0) node[label=above:{\tiny $0$}, color=red] (r) {};
		\draw (.6,-.5) node[label=right:{\tiny $0$}] (v3) {};
		\draw (.8,-1) node[label=below:{\tiny $1$}] (v32) {};
		\draw (r) -- (v3) -- (v32);
	\end{tikzpicture}
\\
BPC 5 &
	\begin{tikzpicture}[scale=1]
		\tikzstyle{every node}=[draw, fill, circle, inner sep=1pt]
		\draw (0,0) node[label=above:{\tiny $0$}, color=red] (r) {};
		\draw (-.6,-.5) node[label=left:{\tiny $0$}] (v1) {};
		\draw (-.8,-1) node[label=below:{\tiny $1$}] (v11) {};
		\draw (-.4,-1) node[label=below:{\tiny $2$}] (v12) {};
		\draw (0,-.5) node[label=left:{\tiny $0$}, color=red] (v2) {};
		\draw (-.2,-1) node[label=below:{\tiny $2$}] (v21) {};
		\draw (.2,-1) node[label=below:{\tiny $2$}] (v22) {};
		\draw (.6,-.5) node[label=right:{\tiny $2$}] (v3) {};
		\draw (.4,-1) node[label=below:{\tiny $1$}] (v31) {};
		\draw (v1) -- (r)
			  (v11) -- (v1) -- (v12)
			  (v21) -- (v2) -- (v22);
		\draw[thick, color=red] (r) -- (v2);
		\draw (r) -- (v3) -- (v31);
	\end{tikzpicture}
	&
	\begin{tikzpicture}[scale=1]
		\tikzstyle{every node}=[draw, fill, circle, inner sep=1pt]
		\draw (0,0) node[label=above:{\tiny $0$}, color=red] (r) {};
		\draw (-.6,-.5) node[label=left:{\tiny $0$}] (v1) {};
		\draw (-.8,-1) node[label=below:{\tiny $1$}] (v11) {};
		\draw (v11) -- (v1) -- (r);
	\end{tikzpicture}
\\
BPC 4 &
	\begin{tikzpicture}[scale=1]
		\tikzstyle{every node}=[draw, fill, circle, inner sep=1pt]
		\draw (0,0) node[label=above:{\tiny $0$}, color=red] (r) {};
		\draw (-.6,-.5) node[label=left:{\tiny $2$}] (v1) {};
		\draw (-.4,-1) node[label=below:{\tiny $2$}] (v12) {};
		\draw (0,-.5) node[label=left:{\tiny $0$}, color=red] (v2) {};
		\draw (-.2,-1) node[label=below:{\tiny $2$}] (v21) {};
		\draw (.2,-1) node[label=below:{\tiny $2$}] (v22) {};
		\draw (.6,-.5) node[label=right:{\tiny $2$}] (v3) {};
		\draw (.4,-1) node[label=below:{\tiny $1$}] (v31) {};
		\draw (r) -- (v1) -- (v12)
			  (v21) -- (v2) -- (v22);
		\draw[thick, color=red] (r) -- (v2);
		\draw (r) -- (v3) -- (v31);
	\end{tikzpicture}
	&
	\begin{tikzpicture}[scale=1]
		\tikzstyle{every node}=[draw, fill, circle, inner sep=1pt]
		\draw (0,0) node[label=above:{\tiny $0$}, color=red] (r) {};
		\draw (0,-.5) node[label=left:{\tiny $0$}, color=red] (v2) {};
		\draw (.2,-1) node[label=below:{\tiny $2$}] (v22) {};
		\draw (v2) -- (v22);
		\draw[thick, color=red] (r) -- (v2);
	\end{tikzpicture}
\\
BPC 3 &
	\begin{tikzpicture}[scale=1]
		\tikzstyle{every node}=[draw, fill, circle, inner sep=1pt]
		\draw (0,0) node[label=above:{\tiny $0$}, color=red] (r) {};
		\draw (-.6,-.5) node[label=left:{\tiny $2$}] (v1) {};
		\draw (-.4,-1) node[label=below:{\tiny $2$}] (v12) {};
		\draw (0,-.5) node[label=left:{\tiny $0$}, color=red] (v2) {};
		\draw (-.2,-1) node[label=below:{\tiny $2$}] (v21) {};
		\draw (.6,-.5) node[label=right:{\tiny $2$}] (v3) {};
		\draw (.4,-1) node[label=below:{\tiny $1$}] (v31) {};
		\draw (r) -- (v1) -- (v12)
			  (v21) -- (v2);
		\draw[thick, color=red] (r) -- (v2);
		\draw (r) -- (v3) -- (v31);
	\end{tikzpicture}
	&
	\begin{tikzpicture}[scale=1]
		\tikzstyle{every node}=[draw, fill, circle, inner sep=1pt]
		\draw (0,0) node[label=above:{\tiny $0$}, color=red] (r) {};
		\draw (0,-.5) node[label=left:{\tiny $0$}, color=red] (v2) {};
		\draw (-.2,-1) node[label=below:{\tiny $2$}] (v21) {};
		\draw (v21) -- (v2);
		\draw[thick, color=red] (r) -- (v2);
	\end{tikzpicture}
\\
BPC 2 &
	\begin{tikzpicture}[scale=1]
		\tikzstyle{every node}=[draw, fill, circle, inner sep=1pt]
		\draw (0,0) node[label=above:{\tiny $0$}, color=red] (r) {};
		\draw (-.6,-.5) node[label=left:{\tiny $2$}] (v1) {};
		\draw (-.4,-1) node[label=below:{\tiny $2$}] (v12) {};
		\draw (.6,-.5) node[label=right:{\tiny $2$}] (v3) {};
		\draw (.4,-1) node[label=below:{\tiny $1$}] (v31) {};
		\draw (r) -- (v1) -- (v12);
		\draw (r) -- (v3) -- (v31);
	\end{tikzpicture}
	&
	\begin{tikzpicture}[scale=1]
		\tikzstyle{every node}=[draw, fill, circle, inner sep=1pt]
		\draw (0,0) node[label=above:{\tiny $0$}, color=red] (r) {};
		\draw (.6,-.5) node[label=right:{\tiny $2$}] (v3) {};
		\draw (.4,-1) node[label=below:{\tiny $1$}] (v31) {};
		\draw (r) -- (v3) -- (v31);
	\end{tikzpicture}
\\
BPC 1 &
	\begin{tikzpicture}[scale=1]
		\tikzstyle{every node}=[draw, fill, circle, inner sep=1pt]
		\draw (0,0) node[label=above:{\tiny $0$}, color=red] (r) {};
		\draw (-.6,-.5) node[label=left:{\tiny $2$}] (v1) {};
		\draw (-.4,-1) node[label=below:{\tiny $2$}] (v12) {};
		\draw (r) -- (v1) -- (v12);
	\end{tikzpicture}
&
	\begin{tikzpicture}[scale=1]
		\tikzstyle{every node}=[draw, fill, circle, inner sep=1pt]
		\draw (0,0) node[label=above:{\tiny $0$}, color=red] (r) {};
		\draw (-.6,-.5) node[label=left:{\tiny $2$}] (v1) {};
		\draw (-.4,-1) node[label=below:{\tiny $2$}] (v12) {};
		\draw (r) -- (v1) -- (v12);
	\end{tikzpicture}
\end{tabular}
\end{center}
\end{Example}

The key theoretical result of the section, the optimality of the backward algorithm, is summarized as follows:

\begin{Theorem}\label{backthm}
	Let $\mathcal{T}$ be a data set and $\mathcal{L_P}$ be the set of all tree-lines with the same starting point $l_0$ for this data set.
	Let $L_n^{b}, \dots, L_{n-k+1}^{b}$ be the last $k$ BPC tree-lines.
	Then, the backward algorithm returns the optimum $(n-k)^{th}$ BPC tree-line, $L_{n-k}^{b}$.
\end{Theorem}

The proof of this theorem is in the Appendix.

\section{Equivalence of PCA and BPCA in Tree Space}\label{equivalence}

A very important aspect of tree space is that, the notion of orthogonality does not exist. In the Euclidean space equivalent of backward PCA, the orthogonality property ensures that the components do not depend on the method used to find them, i.e., the most informative principal component is the same when forward or backward approaches are used. This powerful property of path-independence brings various advantages to the analyst.

In this section, we will prove that the forward and backward approaches are equivalent in the tree space as well when tree-lines are used. This is a surprising result given the lack of any notion of orthogonality. In practice, this result will ensure that the components of backward and forward approaches in binary tree space are comparable.

We will show this equivalence by proving that, for each $1\leq k \leq n$, the $k^{th}$ PC tree-line and the $k^{th}$ BPC tree-line are equal.
An equivalent statement is that their paths are equal: $p_k^f=p_k^b$. Without loss of generality, we will assume that a consistent tie-breaking method is established for both methods in choosing principal components whenever candidate tree-lines have the same sum of weights.
All the proofs can be found in the Appendix.

\begin{Proposition}\label{pro:1}
	Given an integer $1\leq k\leq n$, let $p_1^f,..., p_k^f$ be the paths of the first $k$ principal components yielded by the forward algorithm and $p_n^b,..., p_{k+1}^b$ be the paths of the last $n-k$ principal components yielded by the backward algorithm, then there exist no $i$ and $j$ such that $1\leq i\leq k<j\leq n$ and $p_i^f=p_j^b$.
\end{Proposition}

This proposition motivates the following theorem:

\begin{Theorem}\label{equivalence}
	For each $1\leq k \leq n$ the $k^{th}$ PC tree-line obtained by the forward algorithm is equal to the $k^{th}$ BPC tree-line obtained by the backward algorithm.
\end{Theorem}

This result guarantees the comparability of principal components obtained by either method, enabling the analyst to use them interchangeably depending on which type of analysis is appropriate at the time.

\section{Numerical Analysis}\label{numerical}
In this section we will analyze two different data sets with tree structure. The first data set consists of branching structures of brain arteries belonging to $98$ healthy subjects. An earlier version of this data set was used in Ayd{\i}n et al. (2009) to illustrate the forward tree-line PCA ideas. In that study they have shown that a significant correlation exists between the branching structure of brain arteries and the age of subjects. Later on, $30$ more subjects are added to that data set, and the set went through a data cleaning process described in Ayd{\i}n et al. (2011). In our study we will use this updated data set. 

The second data set describes the organizational structure of a large company. The details of this data set are propriety information, therefore revealing details will be held back. We will investigate the organizational structural differences between business units, and differences between types of departments.

As stated before, we focus on data trees where nodes are distinctly labeled. When constructing a tree data set, labeling of the nodes is crucial since these labels help determine which nodes in a data tree correspond to the nodes in another, and thus shaping the outcome of the whole analysis. The word correspondence is used to refer to this choice. We will handle the correspondence issue separately for each data set we introduce.

\subsection{Brain Artery Data Set}\label{artery}
\subsubsection{Data Description}
The properties of the data set were previously explained in Ayd{\i}n et al. (2009). For the sake of completeness, we will provide a brief summary.

The data is extracted from Magnetic Resonance Angiography (MRA) images of $98$ heathy subjects of both sexes, ranging from 18 to 72. This data can be found at Handle (2008). Aylward and Bullitt (2002) applied a tube tracking algorithm to construct $3D$ images of brain arteries from MRA images. See also Bullitt et al. (2010) for further results on this set.

\begin{figure*}
[ptb]
\begin{center}
\includegraphics[
natheight=1.4in,natwidth=2.1in,height=1.4in,width=2.1in
]%
{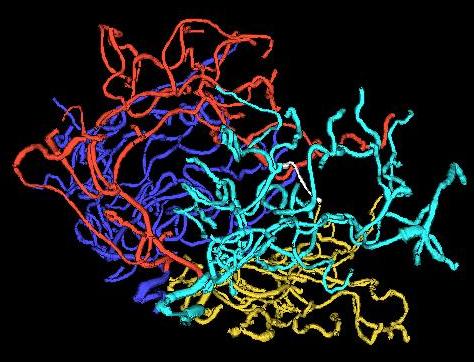}%
\includegraphics[
natheight=1.4in,natwidth=2.1in,height=1.4in,width=2.1in
]%
{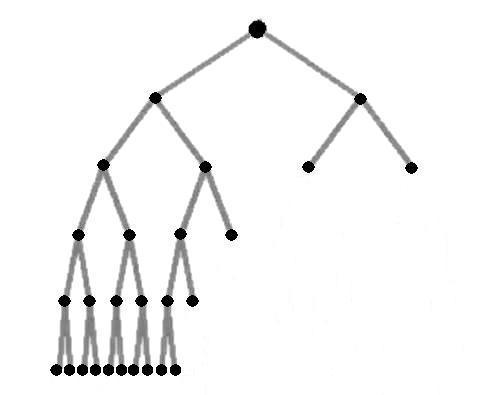}%
\caption{Left panel: Reconstructed set of trees of brain arteries. The colors indicate regions of the brain: Back (gold),
Right (blue), Front (red), Left (cyan). Right panel: An example binary tree obtained from one of the regions. Only branching information is retained.}
\label{3D}%
\end{center}
\end{figure*}

The artery system of the brain consists of $4$ main systems, each feeding a different region of the brain. In Figure \ref{3D} they are indicated by different colors: gold for the back, cyan for the left, blue for the right and red for the front regions. The system feeding each of the regions are represented as binary trees, reduced from the $3D$ visuals seen in Figure \ref{3D}. The reason for this is to focus on the branching structure only. Each node in a binary tree represents a vessel tube between two split points in the $3D$ representation. The two tubes formed by this split become the children nodes of the previous tube. The initial main artery that enters the brain, and feeds the region through its splits, constitutes the root node in the binary tree. The binary tree provided in Figure \ref{3D} (right panel) is an example binary tree extracted from a $3D$ image through this process.

The correspondence issue for this data set is solved as follows. At each split, the child with more number of nodes that descent from it is determined to be the left child, and the other node becomes the right child. This scheme is called descendant correspondence.

The study of brain artery structure is important in understanding how various factors affect this structure, and how they are related to certain diseases. The correlation between aging and branching structure was shown in previous studies (Ayd{\i}n et al. (2009), Bullitt et al. (2010)). The brain vessel structure is known to be affected by hypertension, atherosclerosis, retinal disease of prematurity, and with a variety of hereditary diseases. Furthermore, results of studying this structure may lead to establishing ways to help predict risk of vessel thrombosis and stroke. Another very important implication regards malignant brain tumors. These tumors are known to change and distort the artery structure around them, even at stages where they are too small to be detected by popular imaging techniques. Statistical methods that might differentiate these changes from normal structure may help earlier diagnoses. See Bullitt et al. (2003) and the references therein for detailed medical studies focusing on these subjects.
\subsubsection{Analysis of Artery Data}
The forward tree-line PCA ideas were previously applied to an earlier version of this data set. Our first theoretical contribution of this paper, extension of tree-line PCA to general trees, does not effect this particular data set since all trees in it are binary. Therefore we first focus on the dimension reduction approach we bring. In Ayd{\i}n et al. (2009), only first $10$ principal components were computed, and age effect were presented through first $4$ components. In general, the main philosophy of our dimension reduction or backward  technique is to determine how many dimensions need to be removed for enough noise to get cleared from the data set before the statistical correlations become visible or significant. We ask this question for the brain artery data set and the effect of aging on it, on the updated brain artery data set. Also, Ayd{\i}n et al. (2009) had used the intersection trees as the starting point in calculating the principal components. In this numerical study, we will use the root node as the starting point of the tree-lines.

\begin{figure*}
\begin{center}
\includegraphics[scale=0.5]
{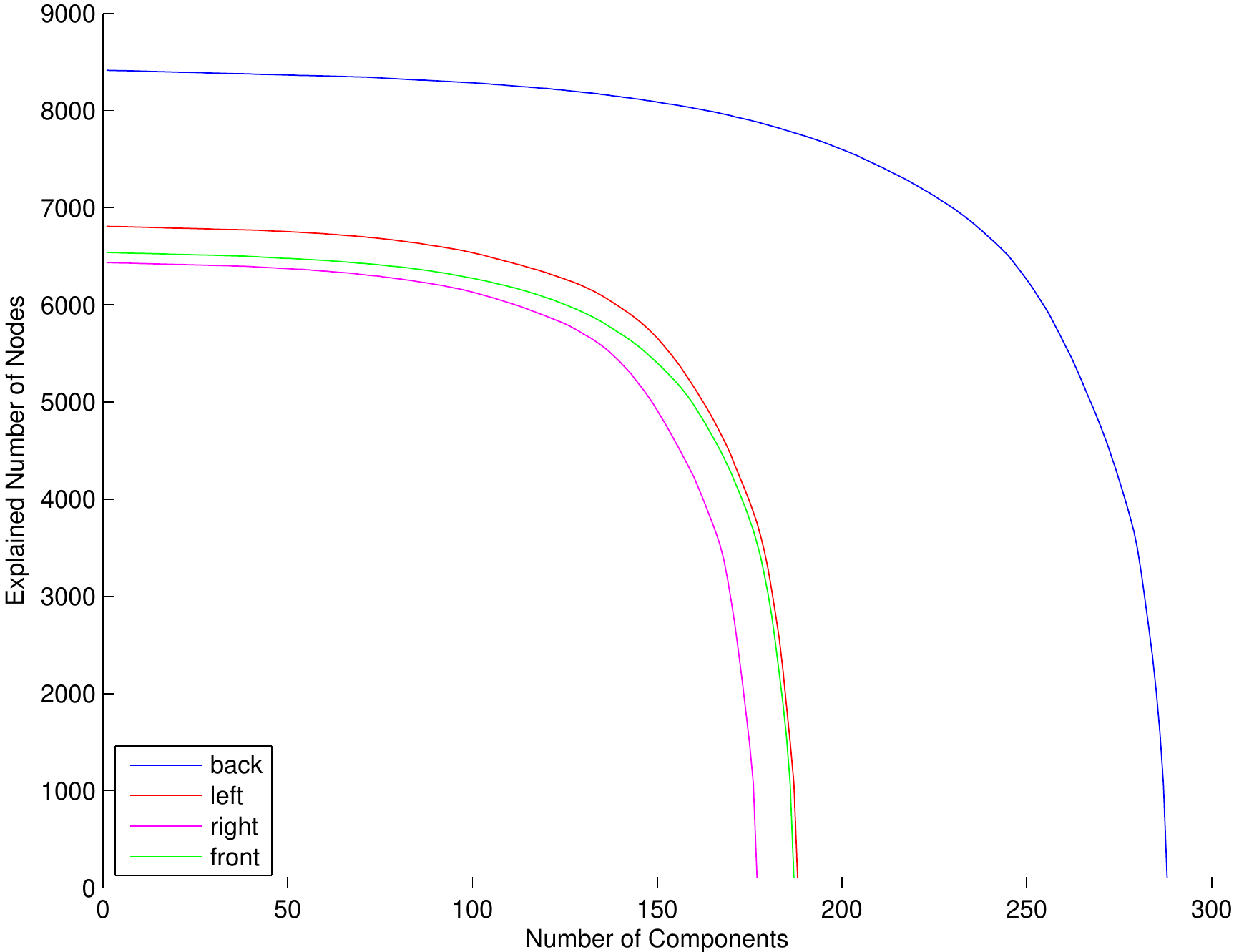}%
\includegraphics[scale=0.5]
{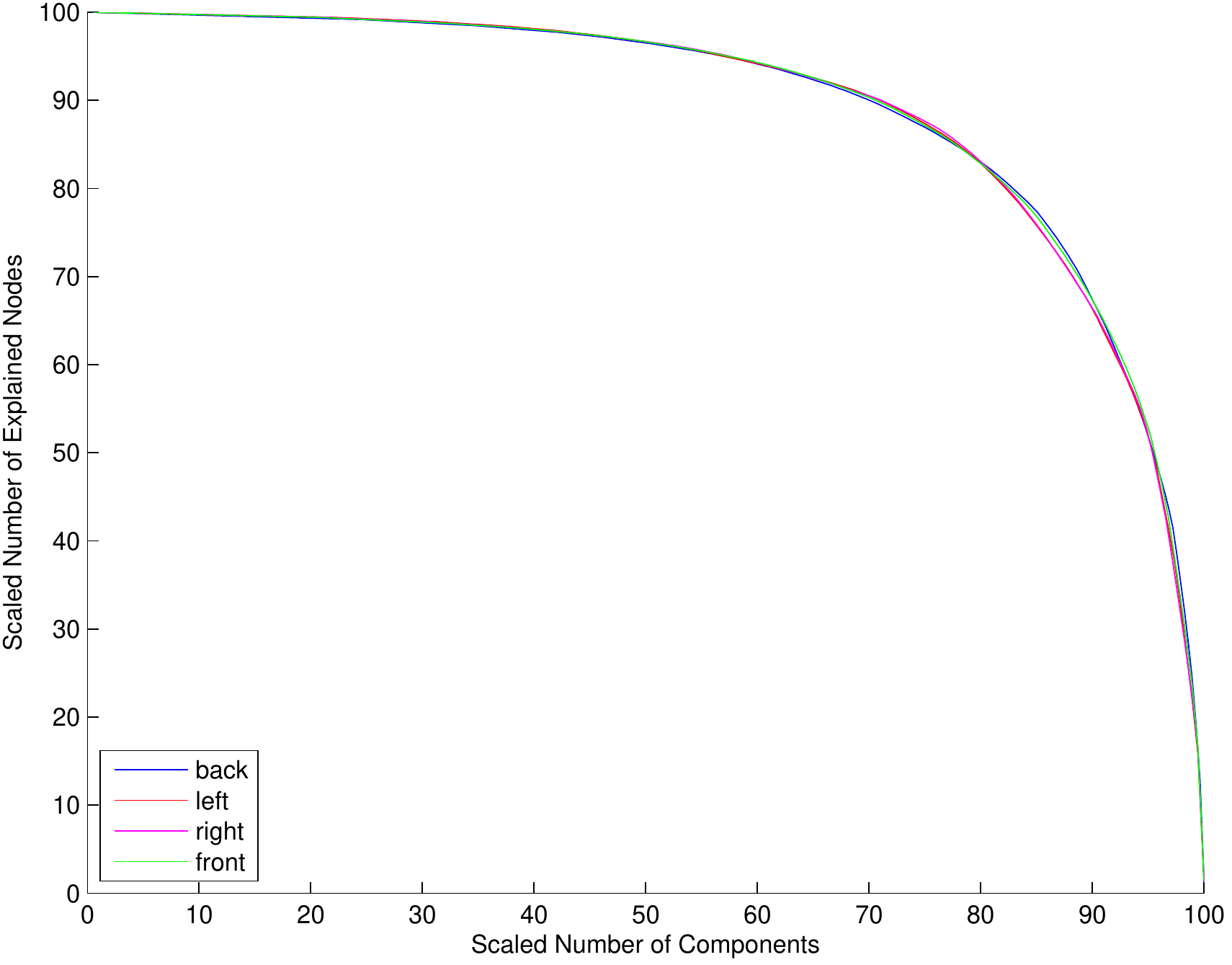}%
\caption{Left panel: X axis represents the total number of backward principal components removed from data. Y axis represents the number of nodes (variation) explained by the remaining subspace after removal. Four subpopulations are shown: Back (blue), Left (red), Right (magenta), Front (green). Right panel: Same information as the left panel is used. For each subpopulation, the total variation and the number of total backward principal components are scaled so that the maximum is $100$.}
\label{Figure2}%
\end{center}
\end{figure*}

An observation on this data set, or any data set consisting of large trees is the abundance of leaves. Many of the leaves of the trees exist in one or few number of data trees. This leads to support trees that are much larger than any of the original data trees. The underlying structures are expected to be seen in upper levels, and most of the leaves can in fact be considered as noise. In our setting, the leaves that only exist in one or few data trees make up the first backward components. A question to ask is, what percentage of variation is created by the low-weight leaves, and what percentage is due to the high-weight nodes, or underlying shape? Figure \ref{Figure2} provides two plots that illustrate an answer.

In Figure \ref{Figure2}, the number of backward components removed from the tree space data is in, versus the total variation explained by the remaining subspace is shown (left panel). The $Y$ values at the $X=0$ point correspond to the total variation before any components are removed. This value is different for each subpopulation, as the sizes of their support trees are different. As backward components are removed from each of the sub-spaces, the variation covered decreases. We can observe that the initial backward components carry very little variation, and therefore result in a very small drop in the total number of explained nodes by the remaining sub-space. This is caused by the very large amount of leaves that aren't part of any underlying structure. The $Y=0$ points for each of the curves mark the total number of principal components that cover the whole data. This number is in fact equal to the number of leaves on the support trees of each of the subpopulations.

On the right panel, we see the same information, only the $X$ and $Y$ axes for each of the curves are scaled so that the maximum corresponds to $100$. The first observation we see in this graph is that, the curves are almost plotted on top of each other: even if the sizes of their support trees are much different, the same percentage of variation is explained by same percentage of principal components in each of these data sets. We can conclude from this that the variation is structured similarly for each of these subpopulations. The second observation is that, the majority of the principal components explain very little variation. In the right panel of Figure \ref{Figure2}, we see that for all the subpopulations, the first $70\%$ of the principal components only cover $10\%$ of the nodes, and the last $10\%$ of these components explain about $70\%$. This data set is known to be very high-dimensional (about $270$ for the back subpopulation). However, Figure \ref{Figure2} shows that a very small ratio of them are actually necessary to preserve the underlying structures.

Our next focus is to see, during the backward elimination process, at which points the age-structure correlation is visible.
\begin{figure*}
\begin{center}
\includegraphics[scale=0.5]{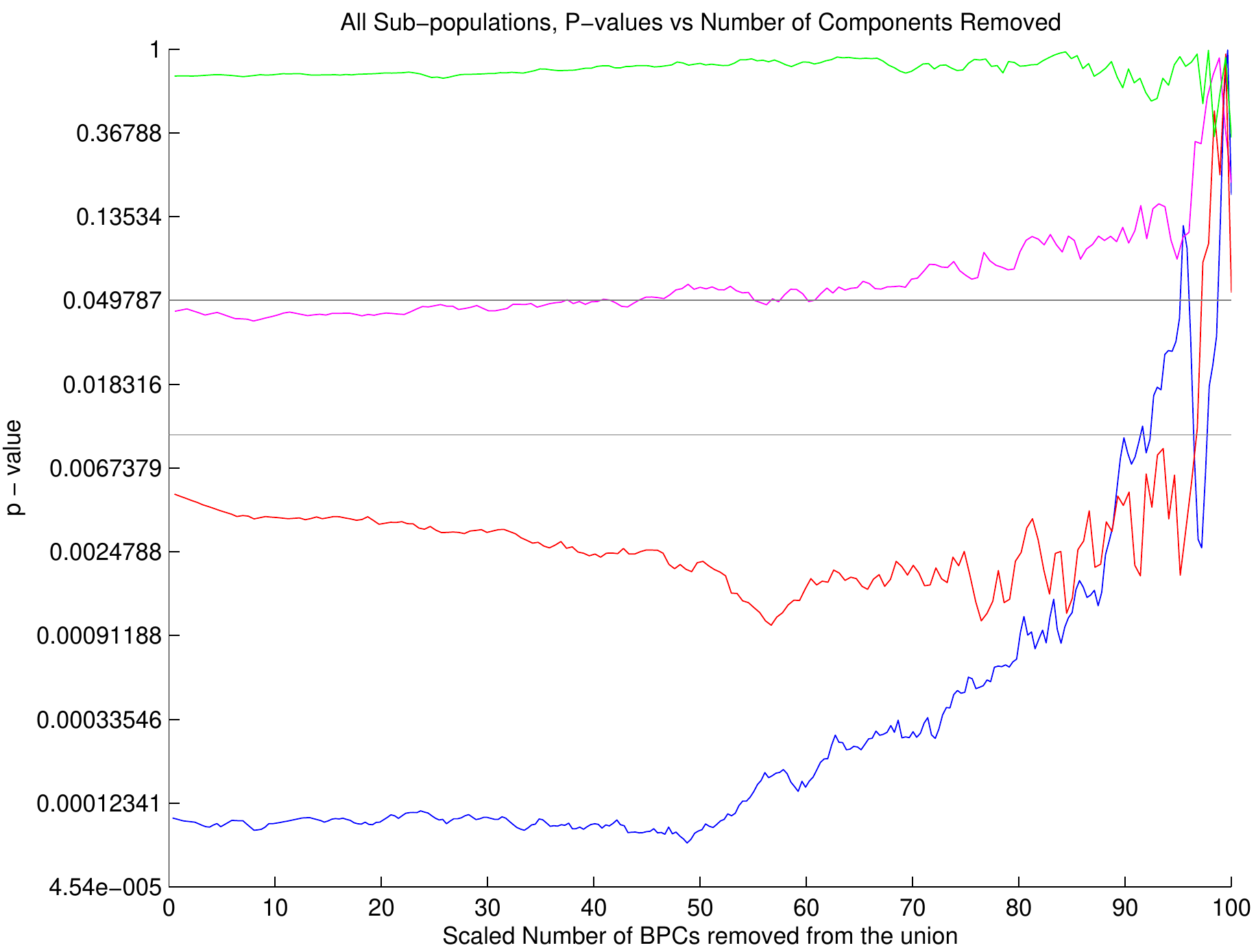}
\caption{$X$ axis represents the scaled number of backward principal components removed from the subspace of each of the subpopulations. At each $X$ value, the data points are projected onto the remaining subspace. The sizes of these projections, plotted against age, show a downward trend (not shown here). Statistical significance of this downward trend is tested by calculating the standard linear regression p-value ($Y$ axis) for the null hypothesis of $0$ slope. $Y$ axis is scaled using natural logarithm, while the $Y$ axis ticks are given in original values. The grey horizontal lines indicate $0.05$ and $0.01$ p-value levels. The subpopulations are colored as:  Back (blue), Left (red), Right (magenta), Front (green). A statistically significant age effect is observed for subpopulations Back, Left and Right.}
\label{Figure3}
\end{center}
\end{figure*}

It was established previously that the branching of brain arteries are reduced with age. Bullitt et al. (2002) noted an observed trend on this phenomenon, while Ayd{\i}n et al. (2009) showed this effect on left subpopulation using principal components. In this paper, for each subpopulation, we start from the whole subspace and reduce it gradually by removing backward principal components. At each step the data trees are projected onto the remaining subspace. The relationship between the age of each data point and the size of the data tree projection is explored by fitting a linear regression line to these two series. These plots are not shown here, but similar ones can be found at Ayd{\i}n et al. (2009). This line tends to show a downward slope, suggesting that the projection sizes are reduced by age. To measure the statistical significance of the observation, the p-values are found for the null hypothesis of $0$ slope. Figure \ref{Figure3} shows the the plots of p-values at each step of removing BPC's, for each subpopulation. The p-values are scaled using natural logarithm while the $Y$ axis ticks are left at their original values. The rule-of-thumb for the p-value is that $0.05$ or less is considered significant. For tight tests, $0.01$ can also be used. Figure \ref{Figure3} provides grey lines for both of these levels for reference.

In Figure \ref{Figure3} we see that, the front subpopulation does not reach the p-value levels that are considered significant at any sub-space. The front region of the brain, unlike the other regions, do not get fed by a direct artery entering the brain from below, but it is fed by vessels extending from other regions. (See Figure \ref{3D}). Therefore it is not surprising that the front vessel subpopulation does not carry a structural property presented by the other three subpopulations.

For other subpopulations, we identify two different kinds of age-structure dependence. First, for left and back subpopulations, the age versus projection size relationship is very sharp until the last $5\%$ of the components are left. Most of the early BPC's correspond to the small artery splits that are abundant in younger population, which people tend to lose as age increases (Bullitt at al. (2002)). Therefore the overall branchyness of the artery trees are reduced. Figure \ref{Figure3} is consistent with this previous observation. The p-value significance gets volatile at the last  $5\%$ of the components, where the BPC's corresponding to the small artery splits are removed, and only the largest components remain in the subspace. These largest components correspond to the main arteries that branch the most. The location-specific relationship between structure and age, noted in Ayd{\i}n et. al. (2009) can be observed for left and back subpopulations towards the end of the $X$ axis. This is the second kind of dependence we observe in the data sets. For right subpopulation, we only observe the first kind, and it does not seem to be as strong as left and back subpopulations. 

Our second focus is to repeat the question of age-structure relationship for the male and female subpopulations. Our data set consists of $49$ male, $47$ female and $2$ trans-gender subjects. We run our analysis for the largest two groups to see how aging effects males and females separately.

\begin{figure*}
\begin{center}
\includegraphics[scale=0.4]{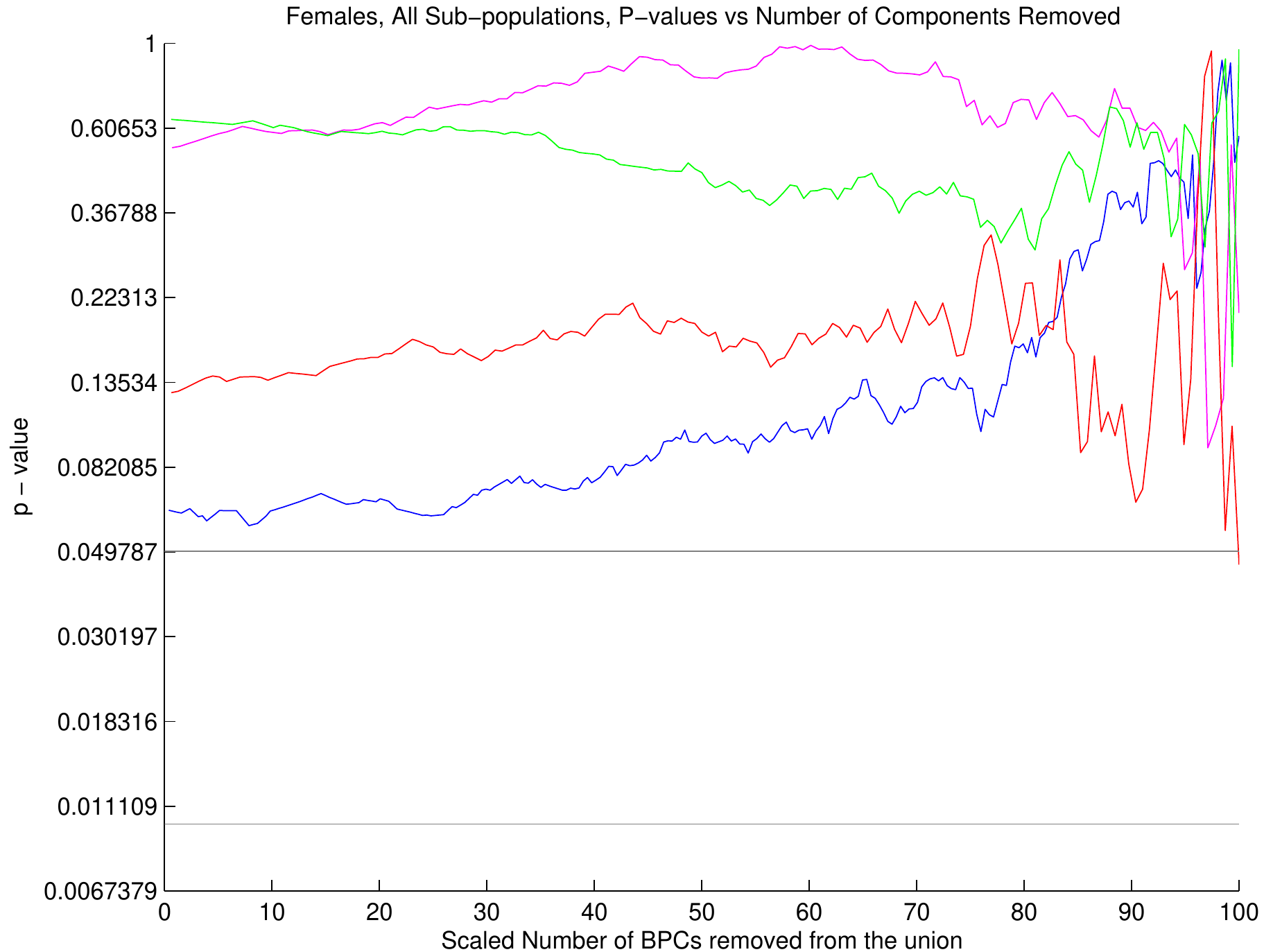}
\includegraphics[scale=0.4]{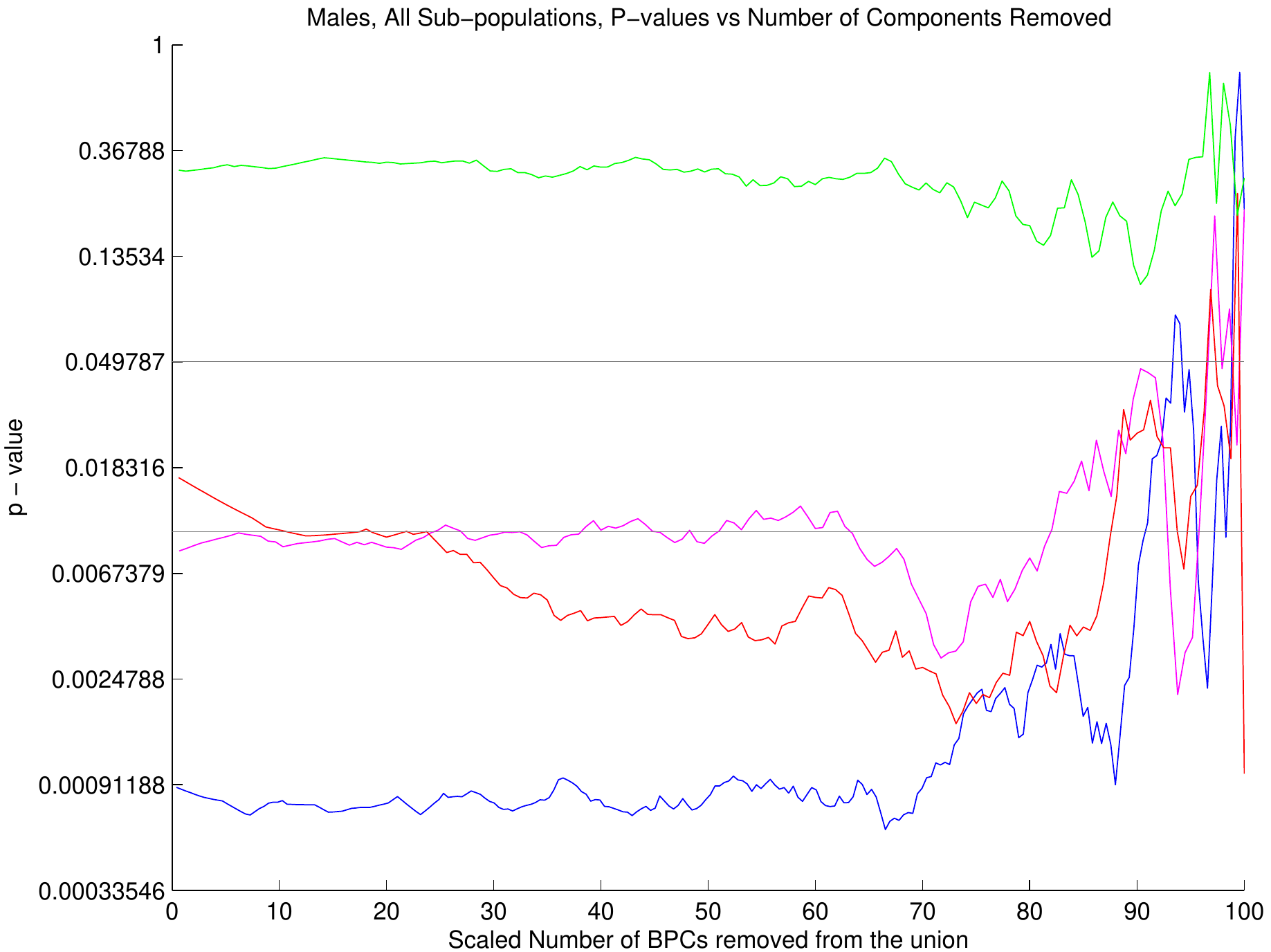}
\caption{The left and right panels are the p-value versus subspace plots for female and male populations. The axes are as explained in Figure \ref{Figure3}. The subpopulations are colored as:  Back (blue), Left (red), Right (magenta), Front (green). For males, a statistically significant age effect is observed for subpopulations Back, Left and Right. No such effect is observed for females.}
\label{Figure4}
\end{center}
\end{figure*}
In Figure \ref{Figure4}, the p-value versus subspace graphs are given for the male and female subpopulations. As before, the front subpopulation does not show any statistical significance at any subspace level. For the other subpopulations, a clear difference between male and female groups emerges.

For the female group, the first kind of structural affect of age (overall branchyness) cannot be observed for any subpopulation. For the location-specific relationship (branchyness of the main arteries) the lowest p-value that could be achieved comes from the right subpopulation at $0.5015$, slightly higher than the rule-of-thumb significance level of $0.05$.

For the male group, the age versus overall branchyness can be observed for left, right and back subpopulations at very significant levels (below $0.01$ p-values). The location-specific relationship can again be observed for these three subpopulations at significant levels.

The study on the full data set implies that two kinds of age-structure relationships can be observed in the whole population using this method. Subsequent analysis of male and female groups shows that the same effects are observed, more strongly, in the male group. Meanwhile, no statistically significant age effect could be observed in the female group using these methods. These results suggest that the brain vessel anatomy of male and females may respond differently to aging: The overall branchyness and the branchyness of longest arteries get reduced by age in males, while these affects aren't apparent for the female group. Therefore the effects observed in the whole population may in fact be driven by the male sub-group.
\subsection{Company Organization Data Set}\label{company}
\subsubsection{Data Description}
In this analysis, we use a company organization data set of a large US company. This data set is a snapshot of the employee list taken sometime during the last ten years. It also includes the information on hierarchical structure and the organizations that employees belong to. The set includes more than two hundred thousand employees active at the time when the snapshot was taken. In this section we will explain the general aspects of the data set that are relevant to our analysis, but we will hold back any specifics due to privacy reasons.

The original company structure can be considered as one giant tree. Each employee is represented as a node. The CEO of the company is the root node. The child-parent relationships are established through the reporting structure: the children of a node are the employees that directly report to that person in the company. Since every employee directly reports to exactly one person (except the CEO, the root node), this system naturally lends itself to a tree representation. A vert important structural property of organization trees is that, each higher-level employee usually has many employees reporting to him/her. Therefore this organization tree is not binary, but a general rooted tree. It has a maximum depth of $13$ levels.

The company operations span various business activities, each main category being pursued by a different business unit of the company. The heads of each of these business units report directly to the CEO. Every person working in the company is assigned to one business unit, and these units form the first level of organization codes. These business units are further divided into sub-organizations, primarily with respect to their geographical locations around the world. A third level of hierarchy again divides these units based on territory and job focus. The last organization level, which we will be using to construct our data sets, is the fourth level of the hierarchy, and is used to define departments that are dedicated to a particular type of job for a particular product or service. For example, the Marketing department responsible of promoting a product group in a given region of one of the business units is an organization at the fourth level of hierarchy. Just like the business unit, every person in the company is assigned to an organization code of second, third and fourth levels. A person working in a particular department shares first, second, third and fourth levels of organization codes with her colleagues working in the same department. 

In this study we will focus on populations of different departments across the company that are assigned to a similar type of job. When the whole organization tree is considered, the directors of these departments are at the fifth level of that tree. To form our data set, we gathered the list of all the directors in the company who are at the fifth level. Then, based on the organization codes, we determined the main job focus of the departments that the directors are leading. We selected four main groups of jobs to compare for our study: finance, marketing, research and development, and sales. The departments that focus on one of these four categories are assigned to those categories. Other departments that focus on different jobs, like legal affairs or IT support, are left out. For each category, each department assigned to that category forms one data point. The director of that department is taken as the root node of the data tree representing the department, and the people who work at that department are nodes of this tree. The structure of the tree is determined by the reporting structure within the department.

The correspondence issue within the data sets requires some attention. A job-based correspondence scheme between two data trees would involve determining which individuals in one department perform a similar function to which individuals at the same reporting level in another department, so that the nodes of those people can be considered "corresponding". With the exception of the directors (who form the root nodes and naturally correspond to each other), this kind of matching is virtually impossible for this data set, since job definitions within one department greatly depends on the particulars of that department's job, and may not match with jobs within another department. Since this job-based correspondence is not possible, we employ the descendant correspondence for the data points. Descendant correspondence was elaborated before for the binary tree setting. In the general tree setting, it works in a similar setting: for the nodes that are the children of the same parent node, the order from left to right is determined by the total number of descendants of each of them. That is, the node with the most number of descendants is assigned as the left-most child, and so on.

\begin{figure*}
\begin{center}
\includegraphics[scale=0.4]{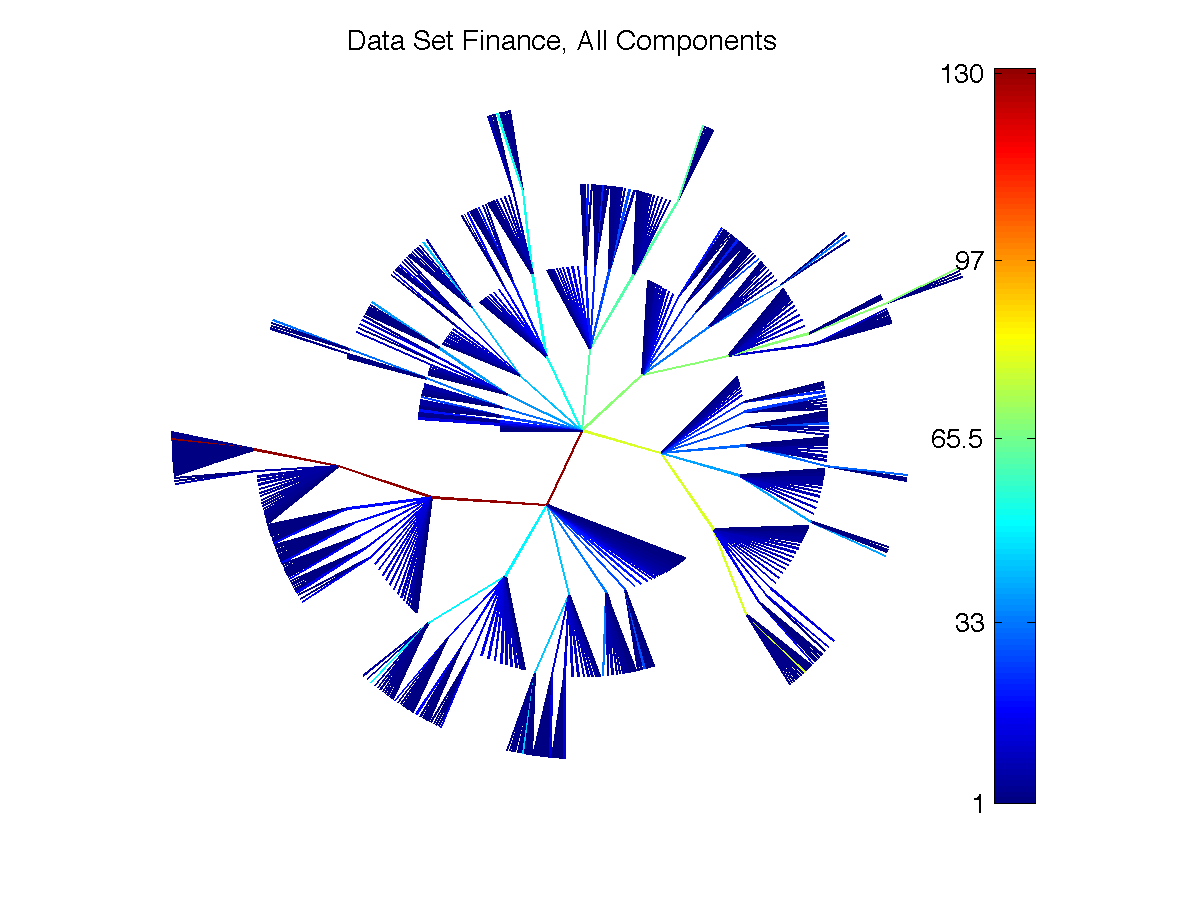}
\includegraphics[scale=0.4]{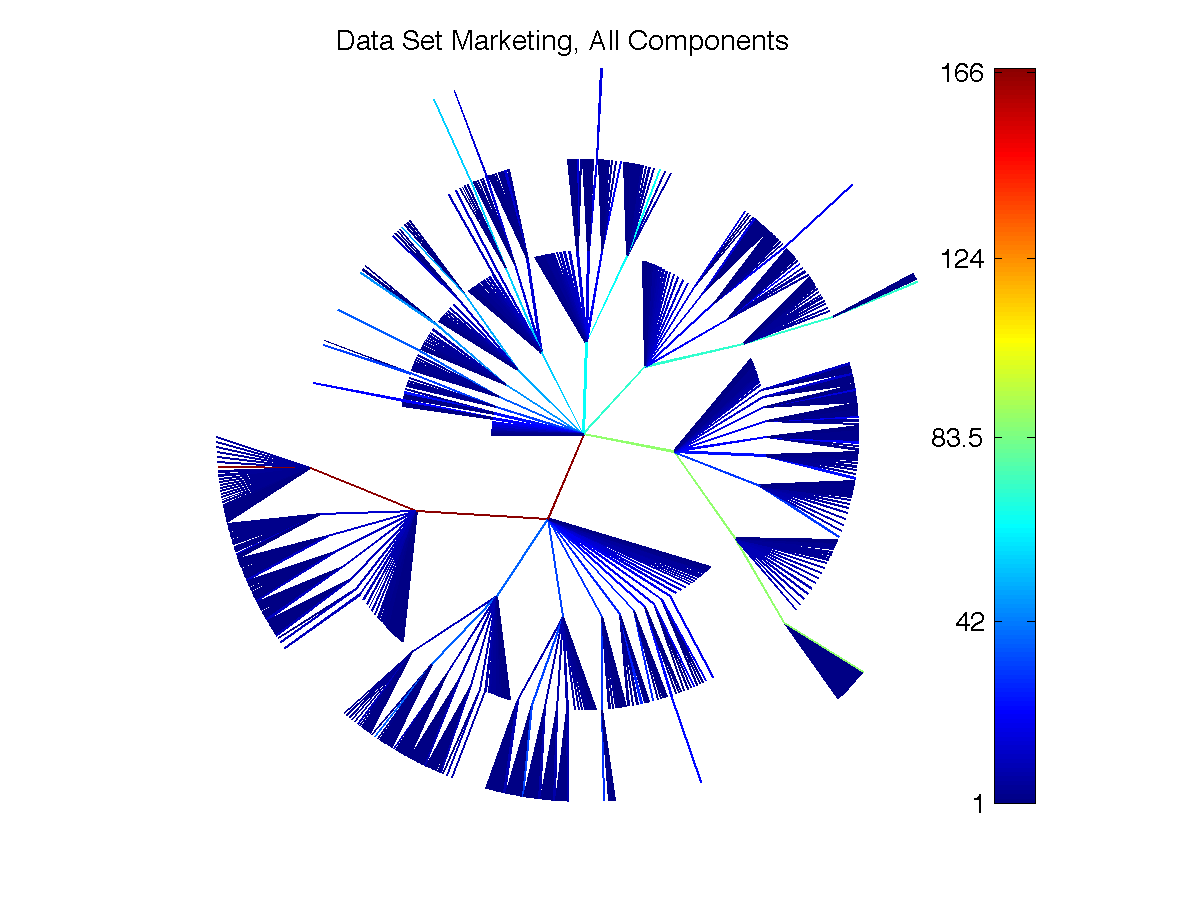}
\includegraphics[scale=0.4]{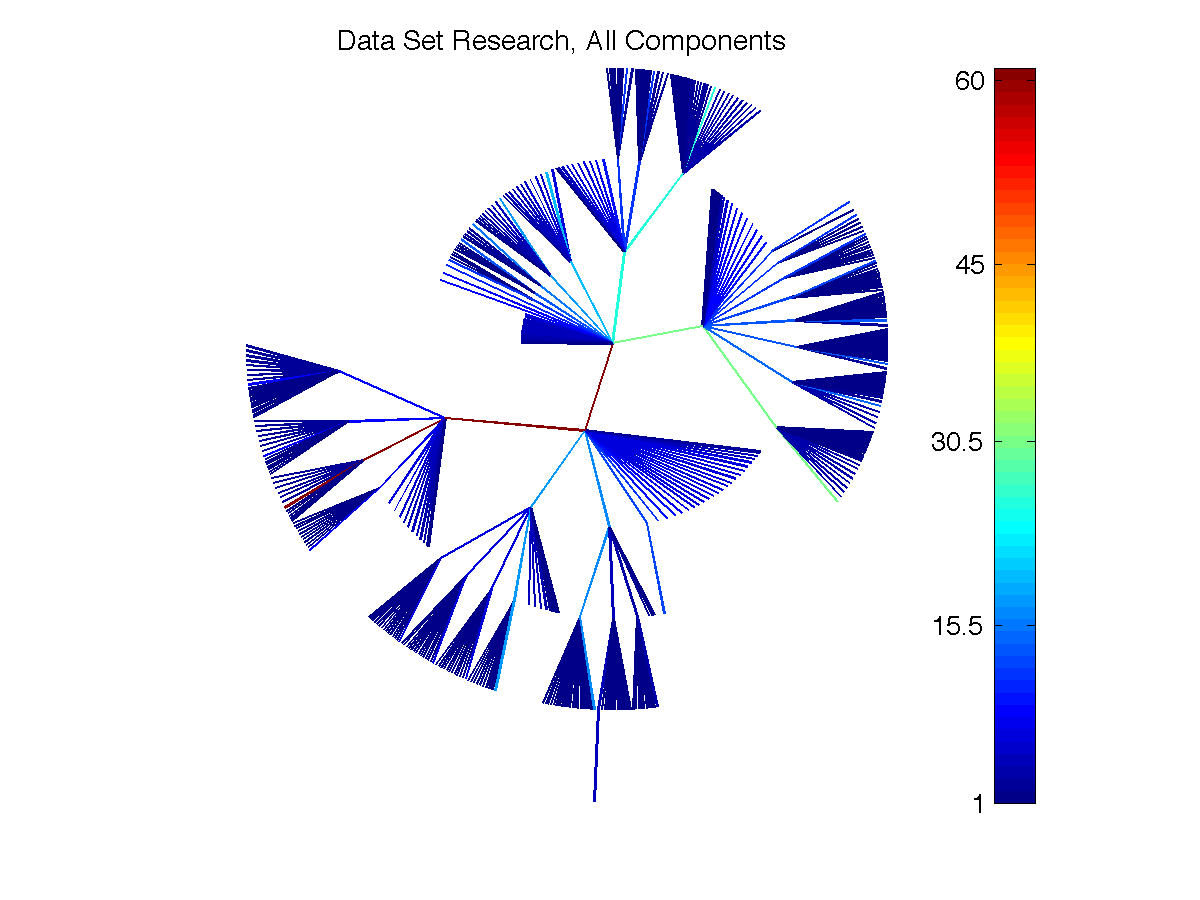}
\includegraphics[scale=0.4]{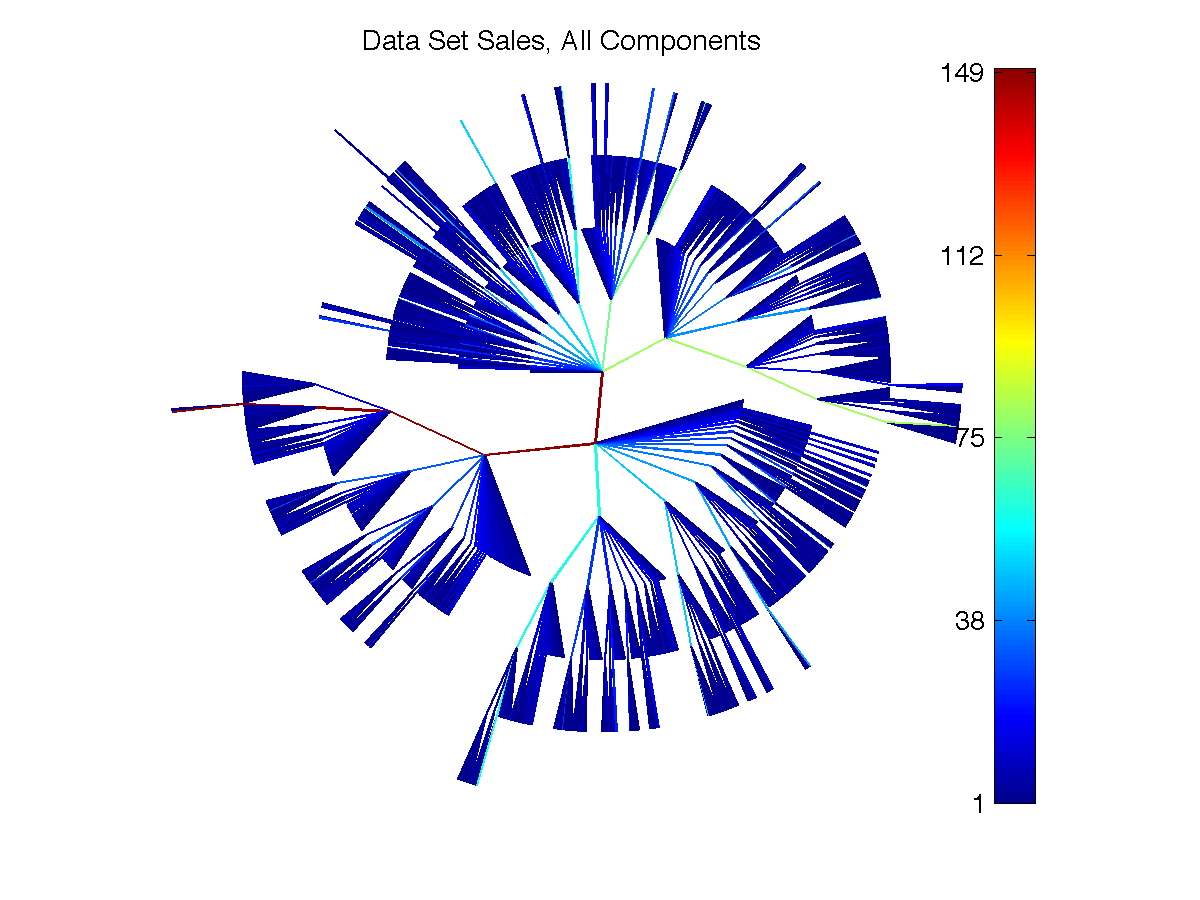}
\caption{Radial drawings of the support trees of four organization subsets: Finance, Marketing, Research and Sales. The root nodes are at the center. The principal components are represented through colors: Earlier BPC's start from the blue end of the color scala while the latter BPC's go towards the red end. Nodes that are in multiple components are colored with respect to the highest total weighted component they are in. The color bar on the right of each panel shows the coloring scheme according to the total weight of each BPC.}
\label{Figure5}
\end{center}
\end{figure*}

The data set of finance departments constructed in this fashion consists of $37$ data trees, with a maximum depth of $6$ levels. The marketing set has $60$ trees, maximum depth of $5$, sales has $41$ trees, maximum depth $5$, and research data set has $20$ trees, maximum depth $6$. The support trees of these sets can be seen in Figure \ref{Figure5}.

Visualizing the organization trees require a somewhat different approach than binary trees. The depth of these trees is not very large: $6$ levels for the deepest data point. However, the node population at each level is very dense. Therefore a radial drawing approach is used to display them. (See Di Battista et al. (1999) for details on this method and many others for graph visualization.) In radial drawing of rooted trees, the root node is at the origin. The root is surrounded by concentric circles centered at the origin. We plot our nodes on these circles, each circle is reserved for the nodes in one level of the tree. The coordinate of each node on a circle is determined by the number of descendants count. For example, for the nodes on the second level, the $360$ degrees available on the circle is distributed to the nodes with respect to the number of descendants they have. Nodes with more descendants get more space. The nodes are put at the middle of the arc on the circle corresponding to the degrees set for that node. The children of that node in the next circle share these degrees according to their own number of descendants. This scheme allows the allocation of most space on the graph to the largest sub-trees and the distribution of nodes on the graph space as evenly as possible.

\subsubsection{Analysis of Company Organization Data}
The comparative structural analysis of these four organization data sets is conducted via the principal component tree-lines. We have run the dimension reduction method for general rooted trees as described in Section \ref{backward}, although the forward method of Section \ref{preliminaries} would have given the same set of components, as shown in Section $4$.

The principal components obtained with this analysis are shown in Figure \ref{Figure5}. They are expressed through the coloring scheme. A color scala starting from dark red, going through shades of yellow, green, cyan and blue and ending at dark blue is used. The components that have higher sum of weights ($\sum{w'(k)}$) are colored with the shades on the red side, and lower sum of weights get the cooler shades. Since the backward principal components are ordered from low sum of weights $\sum{w'(k)}$ to higher, this means the earlier BPC's (lower impact components) are shown in blue, while the stronger components are in yellow to red part of the scala. The color bar on the right of each support tree shows which $\sum{w'(k)}$ corresponds to which shade for that support tree.

The first conclusions on the differences across types of departments come from the comparison of their support tree structure. It can be clearly seen that the sales departments are larger than others in population. Another clear distinction is in the flatness of each organization type. Typically, a flat organization does not have many levels of hierarchy, and most of the workers are do not have subordinates. This is common in organizations of a technical focus. In Figure \ref{Figure5}, we can see that the research departments are visibly flatter than other three types: most of the nodes are at the leaves and not at the interim levels. This is due to the fact that most of the employees in these departments do engineering-research type of work, for which a strongly hierarchical organizational model is  less efficient. The other three data sets, finance, marketing and sales have most of their employees on interim levels, pointing to a strong hierarchy. This seems especially strong in sales departments.

In the next figure (Figure \ref{Figure6}), the effect of reducing the principal components gradually on the amount of nodes explained is shown. This figure is constructed in the same way as Figure \ref{Figure2}, right panel.
\begin{figure*}
\begin{center}
\includegraphics[scale=0.6]{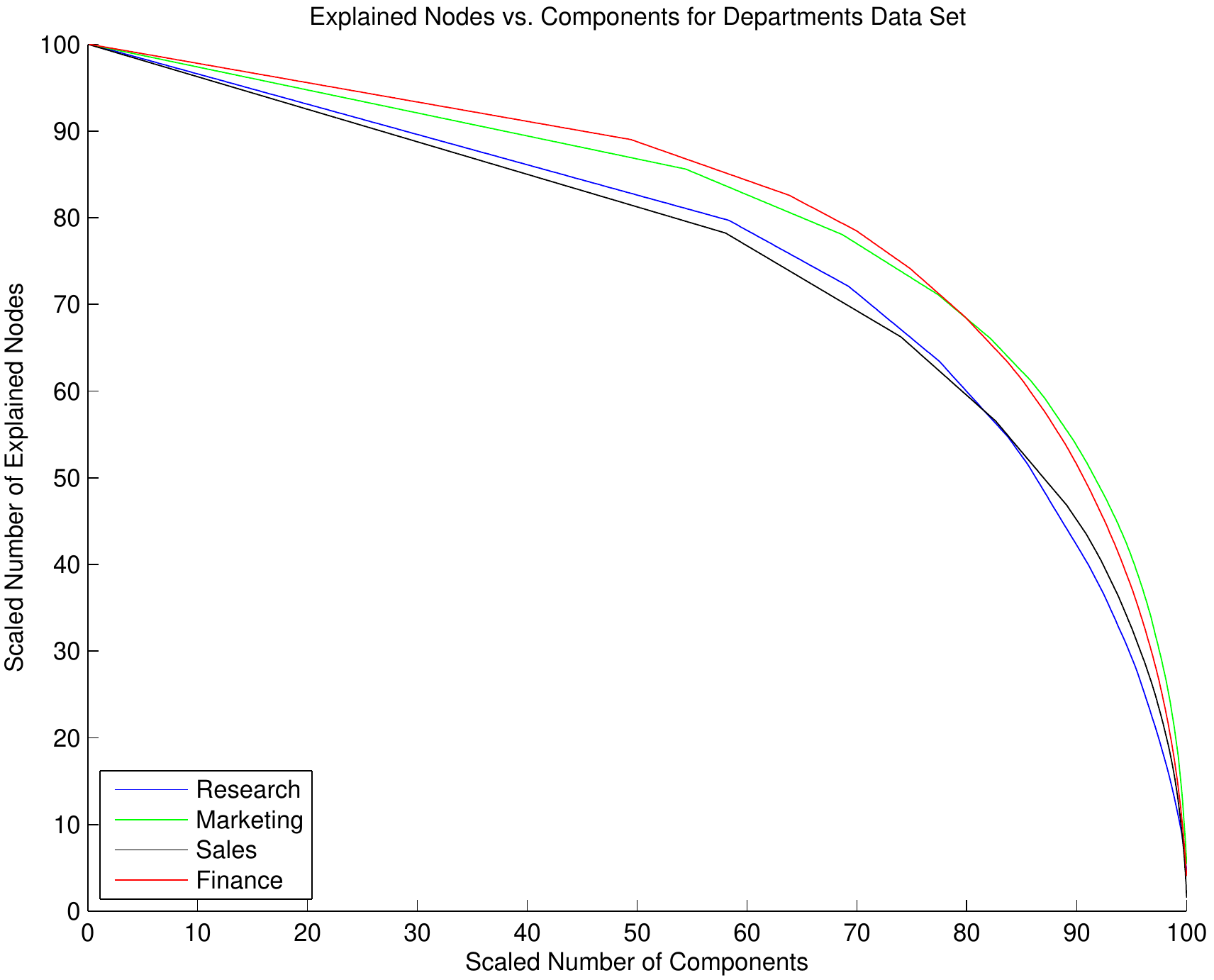}
\caption{The $X$ axis is the number of backward principal components subtracted from the subspace. The $Y$ axis is the amount of nodes that can be explained by the remaining subspace at each $X$ level. Both of the axes scaled within themselves so that the highest $X$ and $Y$ coordinates for all of the organization curves are $100$. The blue curve is for research, green is for marketing, black is for sales and red is for finance.}
\label{Figure6}
\end{center}
\end{figure*}
Figure \ref{Figure6} shows that none of the organization data sets have a very concave variation-versus-components curve like the brain artery set did. Therefore for the organizational structure setting, the earlier BPC's have more potential to carry information compared to the artery setting. Between the organization data sets, we see that the curves belonging to research and sales are very close to each other (the less concave pair), while the curves of finance and marketing are shape-wise close (the more concave pair). The concavity of these curves depend on what percentage of the variation is explained by the early BPC's, and what percentage by the later, stronger components. A very concave curve means that most of the nodes of the data set can in fact be expressed through a small number of principal components. This means that the structures within the data points are not very diverse: the data trees of the set structurally look like each other, allowing a smaller number of PC's to explain more of the nodes. Vice versa, a less concave curve points to a data set where a small portion of the principal components are not enough to explain many nodes due to the diversity in the structures of the data points. Figure \ref{Figure6} shows that finance and marketing departments are more uniformly structured than research and sales departments. I.e., two random finance data trees are more likely to have a shorter distance to each other than two random research data trees.

A variation-versus-components curve is helpful in establishing the trend in the distribution of variation within the data set: the earlier BPC's express nodes that are not common across the data points, and the later BPC's cover the nodes that are common to most data points. The next, and more in-depth question is that, how these more common and less common nodes are distributed among the data points themselves? To answer this question, we divide the set of all BPC's into two subsets. The first $90\%$ of the BPC's on the $X$ axis of Figure \ref{Figure6} form the one set (SET $2$). These BPC's collectively represent the subspace where the less-common-nodes are in. The remaining $10\%$ of the BPC's form the other set (SET $1$). These BPC's express the subspace where the more common structures are in. For any data tree $t$, the projection of it onto SET $1$ ($P_{SET 1}(t)$) represents the portion of the tree that is more common with other data trees in the data set. The projection of $t$ onto SET $2$ ($P_{SET 2}(t)$) carries the nodes of it that are less common with others.  Since these two sets are complementary, the two projections of $t$ would give $t$ itself when combined: $P_{SET 1}(t) \bigcup P_{SET 2}(t) = t$.

\begin{figure*}
\begin{center}
\includegraphics[scale=0.6]{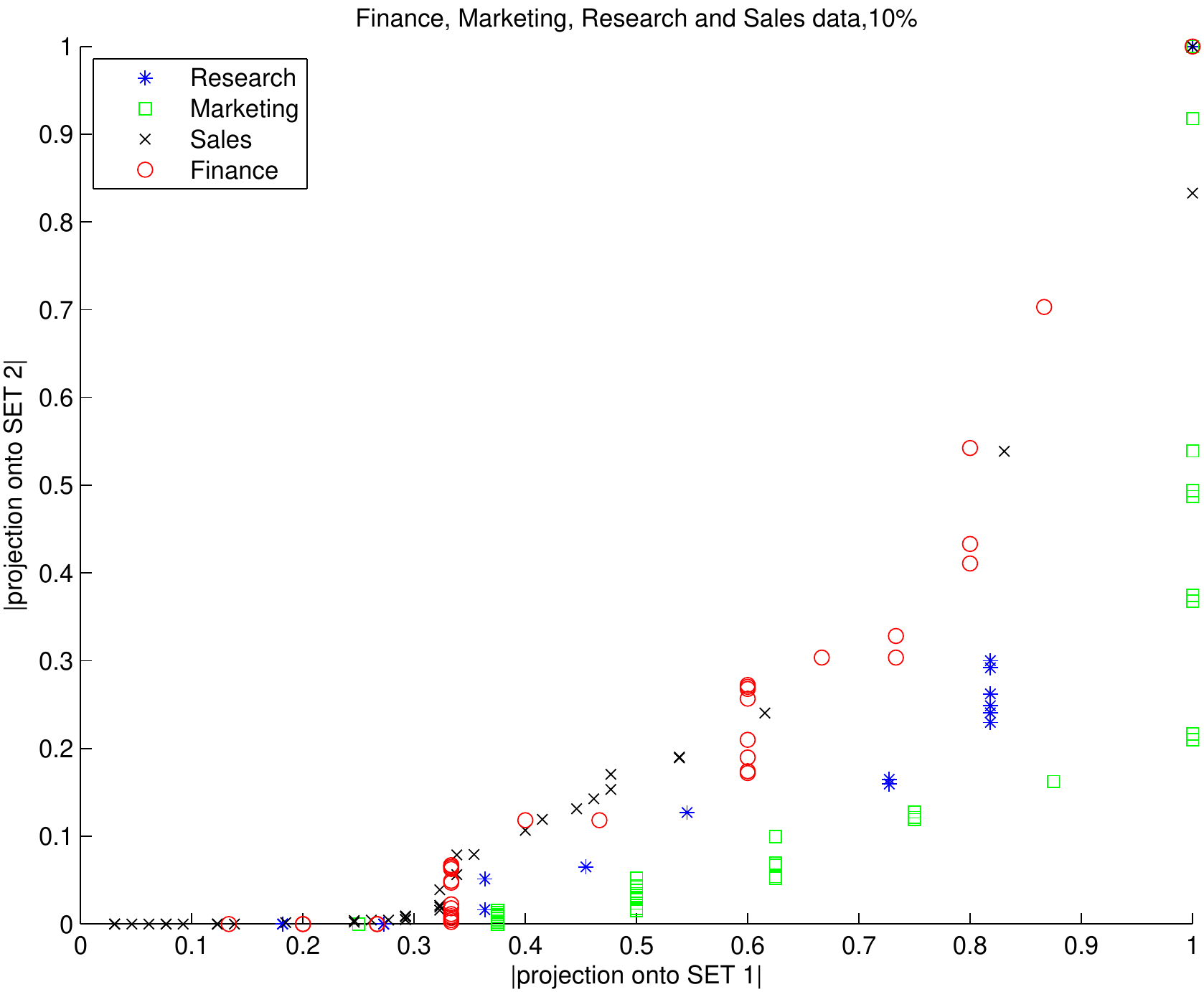}
\caption{The data points of each of the data sets: Research (blue stars), marketing (green squares), sales (black crosses) and finance (red circles). For each of the data points, the length of its projection onto SET $2$ is on the $Y$ axis, and the length of its projection onto SET $1$ is given on the $X$ axis. Each of these axes are scaled such that the highest coordinate for each data set is $1$ on each of the axes.}
\label{Figure7}
\end{center}
\end{figure*}

Figure \ref{Figure7} shows how the nodes in SET $1$ and $2$ are distributed among the data trees for each of the organization data sets. For each data point, the length of its projection onto SET $2$ is on the $Y$ axis, and the length of its projection onto SET $1$ is given on the $X$ axis. Each of these axes are scaled such that the highest coordinate for each data set is $1$ on each of the axes. Blue stars denote the research data points, green squares are marketing data points, black crosses are sales data points and red circles are finance data points.
In Figure \ref{Figure7}, it can be seen that none of the data points are above the $45$ degree line. This is an artifact of the descendant correspondence.

A very interesting aspect of Figure \ref{Figure7} is that, the data points of each data set visually separate from each other. This is especially true for the marketing departments which follow a distinctly more convex pattern compared to other kinds of departments.

For finance departments, we observe an almost linear trend, starting from around $X=0.3$. The bottom left data points are trees that are small in general: they contain little of the common nodes set and almost none of the non-common set. As we go top-right, the trees grow in SET $1$ and SET $2$ spaces proportionally. A similar pattern is there for sales departments, with the exception of a group of data points lying on the $X$ axis, pointing to a group of very small departments that only consist of the main structure nodes. The research departments follow a lower angle pattern. However, this might be due to the one outlier department at the coordinate $(1,1)$, pushing all others to the left/bottom of the graph.

The most significant pattern on this graph belongs to the marketing group. Unlike other departments, there is no linear alignment trend. The set seemingly consists of two kinds of departments: First is the group with very little projection on SET $2$, and varying  sizes of projection on SET $1$. These are relatively small departments. The second is a group of departments that contain all the nodes represented by SET $1$  (therefore the 'common structure' part of the trees are common to all of these trees), and varying, but large amounts of nodes represented in SET $2$. These trees are much larger than the first trees of the group. These two different modes of structure within this group may be due to particular kind of marketing activity, product family, etc they focus on. The details of activities of each department is not part of our data set, therefore we are not able to offer a reason for this separation. Note that two data sets that are shown to be structurally similar in Figure \ref{Figure6}, finance and marketing, are the furthest apart sets in Figure \ref{Figure7}. This is because Figure \ref{Figure6} focuses on the overall dispersion of variation, while Figure \ref{Figure7} focuses on the relative differences between the individual data trees.

\section{Appendix}\label{appendix}

\textbf{Proof of Lemma \ref{lemma1}:}

	Since $l_i = l_{i-1}\cup v_i$, we have that
	\[
		d(t,l_i) =
		\begin{cases}
			d(t,l_{i-1})-1 & \text{if } v_i\in t,\\
			d(t,l_{i-1})+1 & \text{otherwise.}\\
		\end{cases}
	\]
	In other words, the distance of the tree to the line decreases as we keep adding nodes of $p_L$ that are in $t$, and when we step out of $t$, the distance begins to increase. \qed

\textbf{Proof of Lemma \ref{lemma2}:}

	For simplicity, we only prove the statement for $q=2$.
	Assume that
	\[
		L_1=\{l_{1,0}, l_{1,1}, \ldots, l_{1,k_1} \},
		L_2=\{l_{2,0}, l_{2,1}, \ldots, l_{2,k_2} \}
	\]
	with $l_0=l_{1,0}=l_{2,0}$, and
	\begin{eqnarray*}
		l_{1,i}=l_{1,i-1}\cup v_{1,i} & \text{ for } 1\leq i \leq k_1,\\
		l_{2,j}=l_{2,j-1}\cup v_{2,j} & \text{ for } 1\leq j \leq k_2.
	\end{eqnarray*}
	Also assume
	\begin{equation}\label{eqn:1}
		P_{L_1}(t)=l_{1,r_1}
	\end{equation}
	and
	\begin{equation}
		P_{L_2}(t)=l_{2,r_2}.
	\end{equation}
	
	Let $f(i,j)$ be the distance between the trees $t$ and $l_{1,i}\cup l_{2,j}$, for $1\leq i \leq k_1$ and $1\leq j \leq k_2$.
	Using lemma \ref{lemma1}, equation (\ref{eqn:1}) means
	\begin{eqnarray*}
		v_{1,i}\in t, & \text{ if } i\leq r_1 \text{, and }\\
		v_{1,j}\in t, & \text{ if } j\leq r_2.
	\end{eqnarray*}
	Hence,
	\begin{eqnarray}\label{eqn:2}
		f(i,j) \leq f(i-1,j), & \text{ if } i\leq r_1,\\
		f(i,j) \geq f(i-1,j), & \text{ if } i> r_1. \nonumber
	\end{eqnarray}
	By symmetry, we have
	\begin{eqnarray}\label{eqn:3}
		f(i,j) \leq f(i,j-1), & \text{ if } j\leq r_2,\\
		f(i,j) \geq f(i,j-1), & \text{ if } j> r_2. \nonumber
	\end{eqnarray}
	Overall, equations (\ref{eqn:2}) and (\ref{eqn:3}) imply that the function $f$ attains its minimum at $i=r_1, j=r_2$, which is what we had to prove. \qed

\textbf{Proof of Theorem \ref{theorem1}:}

	The definition of $k^{th}$ PC tree-line in terms of paths is equivalent to the equation
	\begin{eqnarray*}
		p_{k}^f & = & \arg \min_{p_L \in \mathcal{P} } \sum_{t\in \mathcal{T}} d\left(t,l_0\cup \left(\left(\cup_{i=1\cdots k-1} p_i^f \cup p_L \right)\cap t\right)\right)\\
			    & = & \arg \min_{p_L \in \mathcal{P} } \sum_{t\in \mathcal{T}} \left| t \setminus \left( l_0\cup \left(\left( \cup_{i=1\cdots k-1} p_i^f \cup p_L \right)\cap t \right)\right)\right | +  \left| \left( l_0\cup \left(\left( \cup_{i=1\cdots k-1} p_i^f \cup p_L \right)\cap t \right)\right) \setminus t\right |\\
			    & = & \arg \min_{p_L \in \mathcal{P} } \sum_{t\in \mathcal{T}} \left| t \setminus \left( l_0\cup p_1^f \cup\cdots \cup p_{k-1}^f \cup p_L \right)\right | +  \left| \left( l_0\cup \left(\left(p_1^f \cup\cdots \cup p_{k-1}^f \cup p_L \right)\cap t \right)\right) \setminus t\right |\\
			    & = & \arg \min_{p_L \in \mathcal{P} } \sum_{t\in \mathcal{T}} \left| t \setminus \left( l_0\cup p_1^f \cup\cdots \cup p_{k-1}^f \cup p_L \right)\right | +  \left| l_0 \setminus t\right |\\
			    & = & \arg \min_{p_L \in \mathcal{P} } \sum_{t\in \mathcal{T}} \left| t \setminus \left( l_0\cup p_1^f \cup\cdots \cup p_{k-1}^f \cup p_L \right)\right |\\
			    & = & \arg \min_{p_L \in \mathcal{P} } \sum_{t\in \mathcal{T}} \left| t \setminus \left( l_0\cup p_1^f \cup\cdots \cup p_{k-1}^f \right)\right | - \left| (t\cap p_L) \setminus \left( l_0\cup p_1^f \cup\cdots \cup p_{k-1}^f \right)\right |\\
			    & = & \arg \min_{p_L \in \mathcal{P} } - \sum_{t\in \mathcal{T}} \left| (t\cap p_L) \setminus \left( l_0\cup p_1^f \cup\cdots \cup p_{k-1}^f \right)\right |\\
			    & = & \arg \max_{p_L \in \mathcal{P} } \sum_{t\in \mathcal{T}} \left| (t\cap p_L) \setminus \left( l_0\cup p_1^f \cup\cdots \cup p_{k-1}^f \right)\right |\\
			    & = & \arg \max_{p_L \in \mathcal{P} } \sum_{v\in p_L} w_k(v).
	\end{eqnarray*}
	The last equation correspond to the path with maximum sum of $w_k$ weights in the support tree. \qed

\textbf{Proof of Theorem \ref{backthm}:}

	The definition of $k^{th}$ BPC tree-line (see Equation \ref{eqn:kBCP}) in terms of paths is equivalent to the equation
	\begin{eqnarray*}
		p_{n-k}^b & = & \arg \min_{p_L \in \mathbf{B} } \sum_{t\in \mathcal{T}} d\left(t,l_0\cup \left(\left({\displaystyle\cup_{p\in{\bf B}\setminus \{ p_L\}}p}\right)\cap t\right)\right),\text{ where }\textbf{B}=\mathcal P\setminus \{p_n^b,\dots , p_{n-k+1}^b\} \\
		& = & \arg \min_{p_L \in \mathbf{B} } \sum_{t\in \mathcal{T}} \left|t\setminus l_0\cup \left(\left({\displaystyle\cup_{p\in{\bf B}\setminus \{ p_L\}}p}\right)\cap t\right)\right|+ \left| l_0\cup \left(\left({\displaystyle\cup_{p\in{\bf B}\setminus \{ p_L\}}p}\right)\cap t\right)\setminus t \right| \\
		& = & \arg \min_{p_L \in \mathbf{B} } \sum_{t\in \mathcal{T}} \left|t\setminus l_0\cup \left(\left({\displaystyle\cup_{p\in{\bf B}\setminus \{ p_L\}}p}\right)\cap t\right)\right|+ \left| \left(l_0\setminus t\right)\cup \left(\left(\left({\displaystyle\cup_{p\in{\bf B}\setminus \{ p_L\}}p}\right)\cap t\right)\setminus t\right) \right| \\
		& = & \arg \min_{p_L \in \mathbf{B} } \sum_{t\in \mathcal{T}} \left|t\setminus l_0\cup \left(\left({\displaystyle\cup_{p\in{\bf B}\setminus \{ p_L\}}p}\right)\cap t\right)\right|+ \left| l_0\setminus t \right| \\
		& = & \arg \min_{p_L \in \mathbf{B} } \sum_{t\in \mathcal{T}} \left|t\setminus l_0\cup \left(\left({\displaystyle\cup_{p\in{\bf B}\setminus \{ p_L\}}p}\right)\cap t\right)\right|\\
		& = & \arg \min_{p_L \in \mathbf{B} } \sum_{t\in \mathcal{T}} \left|t\setminus  l_0\cup \left({\displaystyle\cup_{p\in{\bf B}\setminus \{ p_L\}}p}\right)\right|\\
		& = & \arg \min_{p_L \in \mathbf{B} } \sum_{t\in \mathcal{T}} \left|\left(t\cap p_L\right)\setminus \left( l_0\cup \left({\displaystyle\cup_{p\in{\bf B}\setminus \{ p_L\}}p}\right)\right)\right| +\sum_{t\in \mathcal{T}} \left|\left(t\cap \left(\cup_{p\in \mathcal{P}\setminus {\mathbf B}} p\right)\right)\setminus \left( l_0\cup \left({\displaystyle\cup_{p\in{\bf B}}p}\right)\right)\right|\\
		& = & \arg \min_{p_L \in \mathbf{B} } \sum_{t\in \mathcal{T}} \left|\left(t\cap p_L\right)\setminus \left( l_0\cup \left({\displaystyle\cup_{p\in{\bf B}\setminus \{ p_L\}}p}\right)\right)\right|\\
		& = & \arg \min_{p_L \in \mathbf{B} } \sum_{t\in \mathcal{T}} \sum_{v\in \left(t\cap p_L\right)\setminus \left( l_0\cup \left({\displaystyle\cup_{p\in{\bf B}\setminus \{ p_L\}}p}\right)\right)} 1\\
		& = & \arg \min_{p_L \in \mathbf{B} }  \sum_{v\in p_L} w_k'(v).
	\end{eqnarray*}
	From the last equation the result follows. \qed

\textbf{Proof of Proposition \ref{pro:1}:}

	Suppose there exist $i$ and $j$ with $1\leq i\leq k<j\leq n$ and $p_i^f=p_j^b$.
	Without loss of generality, suppose that $j$ is the largest index where the assumption holds.
	Let $p_L$ denote the path $p_i^f=p_j^b$, and let $B=\{p_n^b,..., p_{j+1}^b \}$.
	Since $1\leq i\leq k<j\leq n$, the set of paths $\mathcal{P}\setminus \{B\}$ contains at least two paths.
	Let $v\in p_L$ be the first node from the leaf to the root that has at least two children in $Supp(\mathcal{P}\setminus \{B\})$.
	There are two possibilities:
	\begin{enumerate}[$I.$]
		\item $v\notin l_0$ {\it i.e.} there is at least one path different of $p_L$ in $\mathcal{P}\setminus \{B\}$ that has $v$ as node or
		\item $v\in l_0$.
	\end{enumerate}
	In both cases, $w_j'(u)=0$ for all $u$ in the path $p_L$ from $v$ to the root.
	
	Consider case $I.$
	Let $p_{L'}\in\mathcal{P}\setminus \{B\}$ be a path different from $p_L$ that contains $v$ in it.
	Let $p_v$ be the path from the root to $v$.
	Since $p_L=p_j^b$
	\begin{equation}\label{eqn:ineq1}
		\sum_{u\in p_L\setminus p_v}w_j'(u)=\sum_{u\in p_L}w_j'(u)\leq\sum_{u\in p_{L'}}w_j'(u)=\sum_{u\in p_{L'}\setminus p_v}w_j'(u).
	\end{equation}
	On the other hand, since $p_L=p_i^f$
	\begin{equation}\label{eqn:ineq2}
		\sum_{u\in p_L}w_i(u)\geq\sum_{u\in p_{L'}}w_i(u).
	\end{equation}
Next, we need to show that following holds:
		\begin{equation}\label{eqn:ineq5}
			\sum_{u\in p_{L'}\setminus p_v}w_j'(u)\leq\sum_{u\in p_{L'}\setminus p_v}w_i(u).
		\end{equation}

		To do this, suppose that $\sum_{u\in p_{L'}\setminus p_v}w_j'(u)>\sum_{u\in p_{L'}\setminus p_v}w_i(u)$.
		It implies that there is at least one node $v'$ that has $w_j'(v')>0$ and $w_i(v')=0$.
		Since $w_i(v')=0$, a path that contains $v'$ and is different of $p_{L'}$ was yielded by the forward algorithm before $p_{L'}$.
		However, this implies that there are at least two paths that has $v'$ as node at step $j$ in the backward algorithm, then $w_j'(v')=0$.
		This gives a contradiction.

	It is straightforward to see
		\begin{equation}\label{eqn:ineq6}
			\sum_{u\in p_L\setminus p_v}w_i(u)\leq\sum_{u\in p_L\setminus p_v}w_j'(u).
		\end{equation}
	Let us suppose that the inequality in (\ref{eqn:ineq1}) is strict, {\it i.e.}
	\begin{equation}\label{eqn:ineq3}
		\sum_{u\in p_L\setminus p_v}w_j'(u)<\sum_{u\in p_{L'}\setminus p_v}w_j'(u).
	\end{equation}
	We have
	\begin{eqnarray*}
		\sum_{u\in p_L}w_i(u) & = & \sum_{u\in p_v}w_i(u)+\sum_{u\in p_L\setminus p_v}w_i(u)\\
		& \leq_{(\ref{eqn:ineq6})} & \sum_{u\in p_v}w_i(u)+\sum_{u\in p_L\setminus p_v}w_j'(u)\\
		& <_{(\ref{eqn:ineq3})} & \sum_{u\in p_v}w_i(u)+\sum_{u\in p_{L'}\setminus p_v}w_j'(u)\\
		& \leq_{(\ref{eqn:ineq5})} & \sum_{u\in p_v}w_i(u)+\sum_{u\in p_{L'}\setminus p_v}w_i(u) = \sum_{u\in p_{L'}}w_i(u)\\
	\end{eqnarray*}
	which is a contradiction to equation (\ref{eqn:ineq2}).
	Therefore, equation (\ref{eqn:ineq1}) has to be an equality, {\it i.e.}
	\begin{equation}\label{eqn:ineq4}
		\sum_{u\in p_L\setminus p_v}w_j'(u)=\sum_{u\in p_{L'}\setminus p_v}w_j'(u).
	\end{equation}
	If one or both equations
	\[
		\sum_{u\in p_{L'}\setminus p_v}w_j'(u)<\sum_{u\in p_{L'}\setminus p_v}w_i(u) \text{ and } \sum_{u\in p_L\setminus p_v}w_i(u)<\sum_{u\in p_L\setminus p_v}w_j'(u),
	\]
	holds, then the result follows in the same way as above.
	Finally, let us suppose
	\[
		\sum_{u\in p_{L'}\setminus p_v}w_j'(u)=\sum_{u\in p_{L'}\setminus p_v}w_i(u) \text{ and } \sum_{u\in p_L\setminus p_v}w_i(u)=\sum_{u\in p_L\setminus p_v}w_j'(u),
	\]
	which implies that
	\[
		\sum_{u\in p_{L'}}w_j'(u)=\sum_{u\in p_L}w_j'(u) \text{ and } \sum_{u\in p_{L'}}w_i(u)=\sum_{u\in p_L}w_i(u).
	\]
	Now, since $p_i^f=p_L$, we have $p_L>p_{L'}$ .
	And, since $p_j^b=p_L$, we have $p_L<p_{L'}$.
	Which is a contradiction.
	
	In the case $II$, where $v\in l_0$, let $v'$ be the last node from the root to the leaf in $p_{L}$ that belongs to $l_0$.
	Take $p_{L'}\in\mathcal{P}\setminus \{B\}$ as a different path of $p_L$, and $v''$ as the last node from the root to the leaf in $p_{L'}$ that belongs to $l_0$.
	Let $p_{v'}$ be the unique path from the root to the node $v'$ and $p_{v''}$ the unique path from the root to the node $v''$.
	Since $p_{v'}$ and $p_{v''}$ are contained in $l_0$, we have
	\[
		\sum_{u\in p_{v'}}w_i(u)=\sum_{u\in p_{v''}}w_i(u)=\sum_{u\in p_{v'}}w_j'(u)=\sum_{u\in p_{v''}}w_j'(u)=0.
	\]
	Since $p_L=p_j^b$
	\begin{equation}\label{eqn:ineq1l0}
		\sum_{u\in p_L}w_j'(u)\leq\sum_{u\in p_{L'}}w_j'(u)
	\end{equation}
	On the other hand, since $p_L=p_i^f$
	\begin{equation}\label{eqn:ineq2l0}
		\sum_{u\in p_L}w_i(u)\geq\sum_{u\in p_{L'}}w_i(u).
	\end{equation}
	Similar to case I, we can see that \ref{eqn:ineq1l0} is an equality.
	This gives a contradiction. \qed

\textbf{Proof of Theorem \ref{equivalence}:}

	By the proposition \ref{pro:1}, we have that at step $n-1$ of the forward algorithm there is no tree-line yielded by the forward algorithm equal to $L_n^b$, then $L_n^b=L_n^f$.
	At the step $n-2$, there is no tree-line yielded by the forward algorithm equal to $L_n^b$ or $L_{n-1}^b$.
	Since $L_n^b=L_n^f$, we have the $L_{n-1}^b=L_{n-1}^f$.
	We continue iteratively until step 1.
	At the end, we will have $L_{k}^b=L_{k}^f$ for all $1\leq k \leq n$. \qed



\begin{thebibliography}{99}
\bibitem{aydin}{Ayd{\i}n B, Pataki G., Wang H., Bullitt E. and Marron J. S. (2009) A Principal Component Analysis For Trees, Annals of Applied Statistics {\bf 4 } vol. 3 1597-1615.}
\bibitem{aydin2}{Ayd{\i}n, B., Pataki, G., Wang, H., Ladha, A., Bullitt, E. and Marron, J.S. (2011) Visualizing the Structure of Large Trees. Electronic Journal of Statistics, 5, 405-420.}
\bibitem{billera}{Billera, L. J., Holmes, S. P. and Vogtmann, K. (2001) Geometry of the space of phylogenetic trees. Adv. in Appl. Math., 27:733-767.}
\bibitem{bullitt}{Aylward, S. and Bullitt, E. (2002) Volume rendering of segmented image objects. IEEE Trans. Medical Imaging, 21:998-1002.}
\bibitem{bullitt3}{Bullitt E., Gerig G., Pizer S.M., Aylward S.R. (2003) Measuring tortuosity of the intracerebral vasculature from MRA images. IEEE-TMI 22:1163-1171}
\bibitem{bullitt2}{Bullitt, E., Zeng, D., Ghosh, A., Aylward, S. R., Lin, W., Marks, B. L., Smith, K. (2010) The Effects of Healthy Aging on Intracerebral Blood Vessels Visualized by Magnetic Resonance Angiography, Neurobiology of Aging, 31(2):290300}
\bibitem{dibattista}{Battista, G.D., Eades, P., and Tamassia, R., Tollis, I.G. (1999) Graph drawing�Algorithms for the visualization of graphs. Prentice Hall, Upper Saddle River, NJ.}
\bibitem{handle}{Handle, http://hdl.handle.net/1926/594 (2008)}
\bibitem{hotelling}{Hotelling, H. (1933) Analysis of a complex of statistical variables into principal components. Journal of Educational Psychology, 24:417-441,498--520.}
\bibitem{jolliffe}{Jolliffe I. (2002) Principal Component Analysis, Second Edition, Springer.}
\bibitem{jung}{Jung S., Liu X., Marron J. S. and Pizer S. M. (2010) Generalized PCA via the backward stepwise approach in image analysis, Brain, Body and Machine, Proceedings on an International Symposium on the Occasion of the 25th Anniversary of the McGill Centre for Intelligent Machines, Montreal, (J. Angeles, et al., eds.), Springer, New York, 111-123}
\bibitem{mardia}{Mardia K. V., Kent J. T. and Bibby J. M. (1973) Multivariate Analysis, Academic Press.}
\bibitem{marron}{Marron, J. S., Jung, S. and Dryden, I. L. (2010) Speculation on the Generality of the Backward Stepwise View of PCA, Proceedings of MIR 2010: 11th ACM SIGMM International Conference on Multimedia Information Retrieval, Association for Computing Machinery, Inc., Danvers, MA, 227-230.}
\bibitem{pearson}{Pearson K. (1901) On Lines and Planes of Closest Fit to Systems of Points in Space, Philosophical Magazine 2 (6): 559--572.}
\bibitem{tschirren}{Tschirren, J., Pal\'{a}gyi, K., Reinhardt, J. M., Hoffman, E. A. and Sonka, M. (2002) Segmentation, skeletonization and branchpoint matching a fully automated quantitative evaluation of human intrathoracic airway trees. Proc. Fifth International Conterence on Medical Image Computing and Computer-Assisted Intervention, Part II. Lecture Notes in Comput. Sci., 2489:12-19.}
\bibitem{wang} {Wang H. and Marron J. S. (2007) Object Oriented Data Analysis: Sets of Trees, Annals of Statistics {\bf 35} 1847--1873.}
\end{thebibliography}
\end{document}